%% file: main.tex
\documentclass[journal]{IEEEtran}
\input{setting}

\title{\hspace{-0.5mm}User Mobility Demands Near-Field Communications in Terahertz Band Wireless Networks Beyond 6G}

\author{Peng Zhang,~\IEEEmembership{Graduate Student Member,~IEEE,} Vitaly Petrov,~\IEEEmembership{Member,~IEEE,} Arjun Singh,~\IEEEmembership{Member,~IEEE,} Emil~Bj{\"o}rnson,~\IEEEmembership{Fellow,~IEEE,} and~Josep M. Jornet,~\IEEEmembership{Fellow,~IEEE}
\thanks{Peng Zhang, Vitaly Petrov, and Emil Bj{\"o}rnson are with the KTH Royal Institute of Technology, Stockholm, Sweden (e-mail: \{pezhang,vitalyp,emilbjo\}@kth.se). Arjun Singh is with the State University of New York Polytechnic Institute, Utica, NY, USA (e-mail: singha8@sunypoly.edu). Josep M. Jornet is with Northeastern University, Boston, MA, USA (e-mail: j.jornet@northeastern.edu). This work has been supported in part by Digital Futures at KTH, Grant 2022–04222 from the Swedish Research Council, and the Swedish Foundation for Strategic Research FFL–9 program.
A shorter version of this work has been presented at the IEEE GLOBECOM~\cite{petrov2023near}.}}

\begin{document}
\maketitle

\begin{abstract}
Near-field propagation is often unavoidable at terahertz (THz) frequencies due to the large apertures needed for sufficient array gain, yet near-field operation complicates practical system design, especially under user mobility. This paper asks whether a \emph{mobile} THz link can remain broadband, achieve the desired high rates and coverage, while operating \emph{exclusively} in the radiative far field. To answer this question, we develop a proof-by-contradiction feasibility framework that jointly enforces (i) a far-field requirement based on the Fraunhofer distance and (ii) a reliability requirement specified by a target SNR at the worst-case link distance. We derive closed-form upper bounds on the far-field-feasible bandwidth for stationary and mobile links. We further incorporate practical misalignment through several UE rotation and mobility scenarios. Numerical results show that stationary THz links \emph{can} remain far-field-only with physically realizable apertures while supporting extremely large bandwidths, whereas practical mobile THz systems \emph{cannot}. In practically relevant mobile THz access settings, the far-field-feasible bandwidth becomes a severe limiting factor: achieving tens-of-GHz targets would require unrealistically high UE transmit power. A cross-band comparison further shows that far-field-only operation is largely attainable at sub-6~GHz and, to a significant extent, at mmWave for moderate bandwidths, while near-field-aware designs become \emph{essential} for mobile THz access.
\end{abstract}


\section{Introduction}
To support data-intensive 6G applications with multi-Gbit/s-to-Tbit/s throughput demands under sub-millisecond latency and high-reliability constraints\cite{rajatheva2020white,wang2023road,petkova2024way,trevlakis2023localization}, terahertz (THz) communications operating roughly in the $0.3$--$10$~THz band have been widely recognized as an technology candidate for beyond-6G systems~\cite{rappaport2019wireless, sarieddeen2019terahertz}. The THz spectrum offers ultra-wide contiguous bandwidths on the order of hundreds of gigahertz \cite{akyildiz2022terahertz,sarieddeen2020next}, which naturally enable Tbps-level data rates without relying only on aggressive spatial multiplexing.

However, the same large-aperture arrays that provide high gain at THz frequencies also give rise to an extended \textit{radiative near-field} region~\cite{balanis2016antenna}. In this regime, classical far-field assumptions break down: the field across the receive aperture is no longer well-approximated by a plane wave, i.e., the wavefront curvature over the (finite-size) Rx array becomes non-negligible. Consequently, the received power can no longer be accurately characterized by a single far-field path-loss factor (e.g., a Friis-type link-budget model) that depends only on the link distance and is separable from the array responses. Canonical far-field beamforming also becomes inefficient, instead spatial focusing and range-dependent design become essential~\cite{lu2024tutorial,cui2022near,petrov2024accurate,zhang20236g}. While operating in the \textit{radiative near-field} enables attractive opportunities (e.g., beam focusing~\cite{zhang2022beam} and spatial selectivity~\cite{cui2022near}), in many practical THz architectures that rely on parametric/sparse channel models and codebook-based beam management (often due to hybrid/analog beamforming with limited RF chains), it also introduces considerable modeling and signal-processing complexity, especially under mobility~\cite{petrov2024accurate,han2021hybrid,cui2022near,liu2023near}.

This tension leads to a fundamental question: \emph{can we design a practical mobile THz communication system that operates \underline{exclusively} in the far field, thereby entirely avoiding near-field-specific complexities, or is near-field operation physically unavoidable?} Addressing this question requires going beyond ``how to communicate in the near field'' and instead characterizing the \emph{feasibility} of remaining far-field-only at THz frequencies under realistic link-budget and mobility constraints.

\subsection{Related Works}
\label{sec:related_works}

\subsubsection{Near-field boundary characterization}
A central prerequisite for near-field (NF) communications is to quantify the separation distance beyond which the far-field (FF) approximation becomes sufficiently inaccurate. 
In classical antenna theory, the radiative NF/FF boundary has been most commonly characterized from a \emph{phase-error} viewpoint~\cite{balanis2016antenna,selvan2017fraunhofer}. 
Specifically, the \textit{Fraunhofer} (\textit{Rayleigh}) criterion defines a threshold distance such that the \emph{maximum} phase variation across the effective aperture (or across different array elements) does not exceed a prescribed tolerance (typically on the order of $\pi/8$)~\cite{balanis2016antenna,selvan2017fraunhofer}. 
Early derivations primarily considered simplified geometries, e.g., a point-like transmitter illuminating a fixed-aperture receiver~\cite{stutzman2012antenna}. 
Then, motivated by electrically large apertures in millimeter wave (mmWave)/sub-THz/THz systems, this phase-based distance characterization has been generalized to practical array configurations, including uniform linear arrays (ULAs) and uniform planar arrays (UPAs), as well as array-to-array links~\cite{lu2023near,petrov2023near,renwang2025applicable,zhang2025impact,zhang2025misalignedTWC}. Beyond phase-based criteria, alternative NF boundary definitions have been proposed from several \emph{performance-oriented} perspectives, including received power~\cite{bjornson2020power}, beamforming gain fidelity~\cite{cui2024near}, spatial multiplexing~\cite{bohagen2009spherical,wang2014tens}, and information-theoretic metrics~\cite{jiang2005spherical}. Recently, criteria accounting for polarization effects or mismatch constraints have also been introduced~\cite{zeng2025revisiting,daei2025near}.

\subsubsection{Near-field communications and signal processing}
With the NF regime increasingly relevant for sub-THz/THz systems, foundational works have provided unified overviews of spherical-wave propagation, range-dependent beamforming, and spatial non-stationarity~\cite{liu2023near,bjornson2021primer}. 
In parallel, LoS MIMO channel models based on uniform and non-uniform spherical waves have been developed to accurately capture large-aperture short-range links~\cite{driessen1999capacity,bohagen2009spherical}.

Building on these models, extensive efforts have been devoted to NF signal processing, including channel estimation, beam management, and beam training. 
NF channel acquisition schemes exploit the inherent angle--range structure and spherical-domain sparsity to control pilot and computational overhead~\cite{cui2022channel,lu2023near}. 
For downlink transmission, NF beamforming emphasizes \emph{spatial focusing} (rather than FF steering) to enhance power concentration and improve interference control~\cite{zhang2022beam,zhang20236g}. 
Efficient NF beam training and tracking further accelerate beam search and mitigate the performance loss caused by mismatches between FF codebooks and NF propagation~\cite{cui2022near,wu2023two}. 
Moreover, NF modeling and design have been integrated with emerging paradigms such as reconfigurable intelligent surfaces (RIS) and holographic MIMO~\cite{singh2023wavefront,bjornson2021primer,wei2023tri}.

\subsubsection{Hybrid near--far field (cross-field / mixed-field) operation}
While many NF studies focus on techniques \emph{within} the NF regime, practical mmWave/THz extra-large (XL) MIMO deployments will often span both NF and FF regions due to heterogeneous user distances, aperture sizes, and network geometries. 
This motivates \emph{cross-field} (or hybrid NF/FF) modeling and design, where spherical- and planar-wave behaviors coexist and must be handled in a unified manner~\cite{liu2025near}. 
For example, cross-field THz ultra-large-array systems have been discussed with unified channel modeling and signal processing solutions bridging NF and FF operation~\cite{han2024cross}. 
At the multiuser/network level, ``mixed-field'' scenarios with simultaneous NF and FF users have been analyzed, revealing new interference behaviors specific to XL-aperture regimes~\cite{zhang2023mixed}. 
From an acquisition perspective, hybrid-field channel estimation frameworks for have been developed to accommodate near--far coexistence and mitigate hybrid-domain mismatches~\cite{yue2024hybrid}.

These developments highlight both the opportunities and the considerable complexities associated with near-field THz operation. Importantly, however, the majority of existing works either (a) assume that the system already operates (at least partly) in the NF regime and then optimize communication \emph{within} that regime, or (b) focus on accommodating near--far coexistence through cross-field processing. In contrast, a more fundamental feasibility question remains largely underexplored: is it actually mandatory to consider near-field operation in mobile THz communication systems? In other words, \textit{\uline{can a mobile THz system with realistic misalignment remain exclusively in the far field under practical constraints?}}
This paper aims to answer this question.

\subsection{Our Novelty and Contributions}
To answer the above feasibility question, we develop an analytical framework that links the \emph{radiative far-field requirement} with a \emph{reliability constraint} under mobility, and use it to establish explicit feasibility limits for mobile THz links equipped with planar arrays.

The main contributions are summarized as follows:

\begin{itemize}
    \item We develop a \emph{proof-by-contradiction framework} to assess whether a \emph{mobile} THz link can remain \emph{exclusively} in the far field under practical constraints. The framework jointly enforces: (i) a radiative far-field requirement and (ii) a link-budget requirement, evaluated through a target SNR at the maximum link distance (e.g., at the cell edge).

    \item Within the above framework, we derive \emph{closed-form limits on the maximum feasible bandwidth} (equivalently, the required transmit power) for far-field-only operation in representative planar-array misalignment scenarios. Specifically, we analyze one- and two-angle rotations at the UE side, as well as effective one- and two-angle rotations at the AP side. 

    \item We provide extensive \emph{numerical results to quantify the derived feasibility limits} by plotting the maximum feasible bandwidth and the required transmit power over representative system parameters and misalignment configurations. The results show how mobility and rotation affect the achievable bandwidth and link-budget requirements, offering practical engineering insights for mobile THz system design.

    \item  We utilize the developed framework to \emph{compare mobile operation across different carrier bands} (sub-6\,GHz, mmWave, and THz). The comparison shows that, for typical beyond-6G targets, mobility-induced constraints are markedly more stringent at THz, which makes near-field operation (at least partially) effectively unavoidable. In contrast, far-field-only operation remains substantially more attainable at lower-frequency bands.
\end{itemize}

A shorter version of this work appeared in~\cite{petrov2023near}. This paper substantially extends~\cite{petrov2023near} by, among other improvements: (i)~incorporating practical array misalignment into the analytical framework; (ii)~establishing closed-form far-field feasibility limits for mobile THz links in different deployment configurations; and (iii)~presenting an illustrative comparison of the severity of the studied effects in different candidate frequency bands considered for beyond-6G wireless systems.

\subsection{Organization}
The rest of this paper is organized as follows. Section~II introduces the system model and the generalized planar array gain model. Section~III develops the proof-by-contradiction framework and derives far-field feasibility limits for stationary and mobile links, including explicit results under representative misalignment scenarios. Section~IV presents numerical evaluations and discusses the resulting design insights. Section~V provides the main conclusions of the paper.

\section{System Model}\label{sec:system_model}

\begin{figure}[!t]
    \centering
    \includegraphics[width=0.95\linewidth]{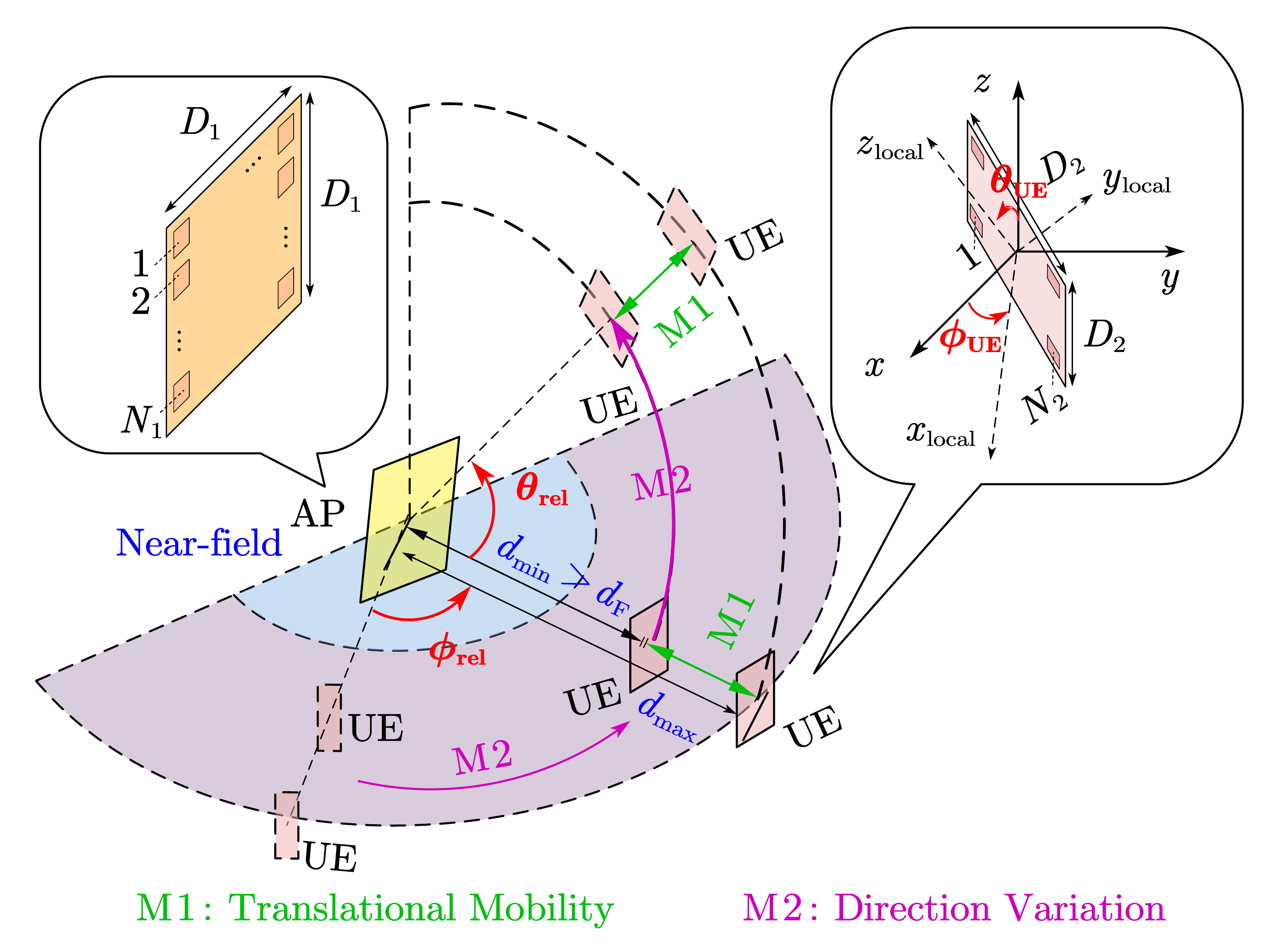}
    \caption{Non-stationary THz link operating exclusively in the far field.}
    \label{fig:system_model}
\end{figure}

In this paper, we consider a point-to-point \emph{non-stationary} THz communication link between a stationary THz access point (THz-AP) and a mobile THz user equipment (THz-UE), as illustrated in Fig.~\ref{fig:system_model}. 
Both nodes are equipped with square planar antenna arrays with half-wavelength inter-element spacing, i.e., $\lambda/2$, where $\lambda=c/f$ is the wavelength corresponding to the carrier frequency $f$. 
The THz-AP employs an $N_1 \times N_1$ array with physical aperture size $D_1 \times D_1$ (m$^2$), while the THz-UE employs an $N_2 \times N_2$ array with aperture size $D_2 \times D_2$ (m$^2$), where $D_i \approx N_i \lambda/2$ for $i\in\{1,2\}$.

Fig.~\ref{fig:system_model} provides a unified geometric illustration that encompasses the different mobility and orientation mechanisms considered in this work. First, \emph{translational mobility} is modeled by an uncertain link distance $d$, measured between the centers of the AP and UE arrays, i.e., $d \in [d_{\min}, d_{\max}]$, where $d_{\min}$ and $d_{\max}$ denote the minimum and maximum separation distances over the considered mobility range. Second, \emph{rotational mobility} captures orientation variations caused by user/device handling and motion. Specifically, we distinguish between (i) UE \emph{rotation}, which directly changes the orientation of the UE array boresight, and (ii) UE translational motion over a spherical surface, which induces \emph{relative angular variations} with respect to the fixed AP boresight. These effects are parameterized by one- and two-angle representations, denoted by $\theta_{\mathrm{UE}}$ and $(\theta_{\mathrm{UE}},\phi_{\mathrm{UE}})$ for UE rotation, and by $\theta_{\mathrm{rel}}$ and $(\theta_{\mathrm{rel}},\phi_{\mathrm{rel}})$ for the induced relative angular deviations, respectively\footnote{Although our primary focus is on terrestrial mobile UEs, the adopted geometry is intentionally three-dimensional and also covers elevated nodes (e.g., UAVs or airborne platforms) as well as indoor multi-floor deployments. 
Including such 3D mobility allows us to assess far-field-only feasibility under a broader set of (possibly extreme) relative orientations and motion patterns, which could strengthen the generality of the analysis derived in Section~\ref{sec:analysis}.}.

We focus on LoS-dominated THz links and assume that beam pointing (or pre-steering) tracks the instantaneous LoS direction so that the main lobe is directed toward the intended node \footnote{For clarity, we consider an idealized digital-array implementation. We hence do not model additional architecture-dependent bandwidth limits, such as those imposed by beam squint in phase-shifter-based analog/hybrid arrays, practical antenna/hardware bandwidth constraints, etc. These effects may further limit the usable bandwidth in practical deployments. Still, the key objective of this paper is to understand whether far-field-only mobile THz operation is feasible even under otherwise favorable assumptions.}. The transmit power $P_{\mathrm{t}}$ is assumed to be uniformly distributed over the signal bandwidth $B$, while the receiver experiences thermal noise and an additional noise figure $N_{\mathrm{F}}$. The same model applies to both downlink and uplink by interpreting $P_{\mathrm{t}}$ and $N_{\mathrm{F}}$ as the corresponding transmitter power and receiver noise figure, respectively. For notational convenience, we also introduce two dimensionless coefficients: the aperture-inequality coefficient $L \triangleq D_1/D_2$ and the mobility coefficient $M \triangleq d_{\max}/d_{\min}$. 

\textit{Planar array gain model}: To accommodate both aligned and misaligned configurations in a unified manner, we next adopt a generalized planar array gain model that separates the element radiation characteristics from the array-level scaling, and we establish the corresponding array-gain expression for planar arrays under arbitrary orientation parameters.

We denote by $f(\theta,\phi)$ the radiation (or field) pattern of an individual antenna element, where $(\theta,\phi)$ represent two generic angular parameters that characterize the direction in space. The normalized power pattern is defined as
\begin{equation}
U_{\mathrm{norm}}(\theta,\phi)
=
\frac{U_{\mathrm{rad}}(\theta,\phi)}{U_{\mathrm{rad,max}}}
=
|f(\theta,\phi)|^{2},
\end{equation}
where $U_{\mathrm{rad,max}}$ denotes the maximum radiant intensity of the element over all directions. Therefore, the corresponding element gain is
\begin{equation}
G_{e}(\theta,\phi) = G_0 |f(\theta,\phi)|^{2},
\end{equation}
where
\begin{equation}
G_0=\left(\int\!\!\!\int \frac{1}{4\pi} |f(\theta,\phi)|^{2} \, d\Omega\right)^{-1},
\label{equ:G_0_definition}
\end{equation}
denotes the element-pattern normalization constant.

Assuming the LoS direction coincides with the boresight, i.e., $(\theta,\phi)=(0,0)$, we have $|f(0,0)|^{2}=1$, whereas any misalignment produces a gain penalty that depends on $|f(\theta,\phi)|^{2}$.

For an $N_1 \times N_1$ planar array at the AP and an $N_2 \times N_2$ planar array at the UE, assuming ideal coherent combining and approximately uniform received power across each array aperture (i.e., beyond the uniform-power distance~\cite{lu2024tutorial}), the corresponding directional antenna gain toward direction $(\theta,\phi)$ at the AP is
\begin{equation}
G_{\mathrm{AP}}(\theta,\phi)
=
N_1^{2} G_{e}(\theta,\phi)
=
 N_1^{2}G_0  |f(\theta,\phi)|^{2},
 \label{eq:G_AP_general}
\end{equation}
and, similarly, the directional antenna gain of UE is
\begin{equation}
G_{\mathrm{UE}}(\theta,\phi)
=
N_2^{2} G_{e}(\theta,\phi)
=
  N_2^{2}G_0 |f(\theta,\phi)|^{2}.
\label{eq:G_UE_general}
\end{equation}

\section{Analysis: Proof by Contradiction}
\label{sec:analysis}

In this section, we develop a \textit{proof-by-contradiction} feasibility framework for mobile THz links. The key idea is to \textit{\textbf{assume far-field-only operation}} and then evaluate whether such an assumption can simultaneously satisfy both the geometric far-field condition and the link-budget (SNR) requirement \textit{under practical transmit-power (and bandwidth) constraints}. Whenever these two constraints are incompatible, the contradiction implies that near-field THz operation is unavoidable. 

The analysis in this section is organized as follows. Section~\ref{sec:scenario0} presents the baseline aligned configuration without rotation and formalizes the proof-by-contradiction framework by jointly imposing (i) the geometric far-field condition and (ii) a worst-case SNR requirement at the maximum link distance. Sections~\ref{sec:scenario1} and~\ref{sec:scenario2} then incorporate UE \emph{rotation} with one and two angles, respectively, and derive the corresponding far-field-only feasibility limits. Finally, Sections~\ref{sec:scenario3} and~\ref{sec:scenario4} consider \emph{relative angular variation} with one and two angles, respectively, which captures motion-induced misalignment with respect to the fixed AP boresight.

\begin{figure}[!t]
    \centering
    \includegraphics[width=0.95\linewidth]{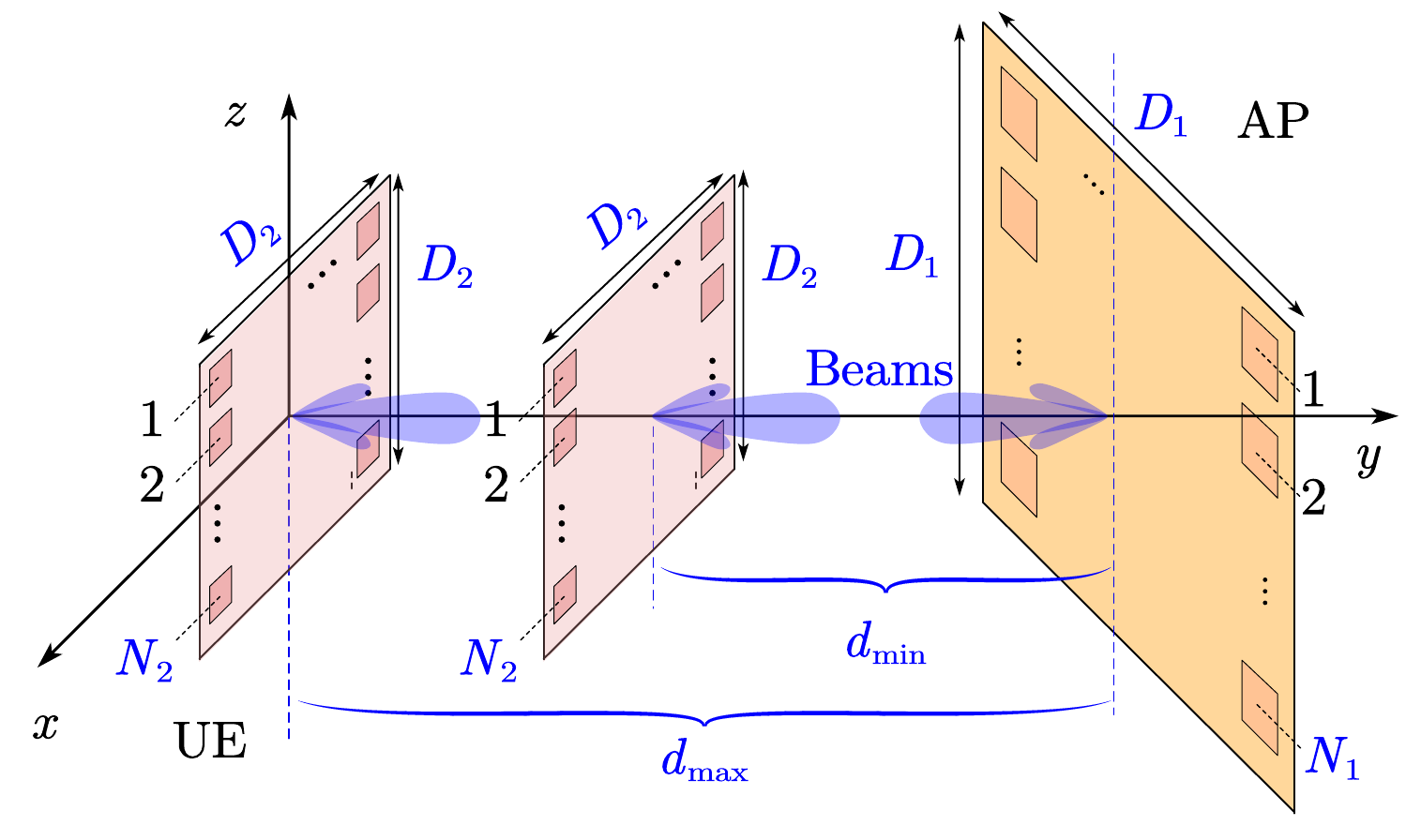}
    \caption{Baseline aligned scenario, where the UE is fixed.}
    \label{fig:scenario0}
\end{figure}

\subsection{Scenario 0: General Framework and Baseline Aligned Case (No Rotation)}
\label{sec:scenario0}

We begin with the baseline stationary scenario, illustrated in Fig.~\ref{fig:scenario0}, where both the AP and UE are perfectly aligned with the LoS direction. 

\textbf{\textit{Condition 1 (Geometric Far-Field Condition).}}
For a prospective THz communication system with a transmitter and receiver equipped with $D_{1}\times D_{1}$ and $D_{2}\times D_{2}$ square arrays, respectively, the Fraunhofer distance $d_{\mathrm{F}}$~\cite{balanis2016antenna,petrov2023near} is given by
\begin{equation}
d_{\mathrm{F}} = \frac{4(D_1 + D_2)^2}{\lambda},
\label{eq:dF_general}
\end{equation}
and the minimum communication distance must satisfy
\begin{equation}
d_{\min} \geq d_{\mathrm{F}},
\label{eq:cond1}
\end{equation}
which ensures the propagation operates in the far-field~region.

\textbf{\textit{Condition 2 (SNR Requirement).}} Let $\mathsf{SNR}(d)$ denote the received SNR at distance $d$. The system must satisfy
\begin{equation}
\mathsf{SNR}(d_{\max}) \geq \mathsf{SNR}_{\mathrm{th}},
\label{eq:cond2_base}
\end{equation}
where $\mathsf{SNR}_{\mathrm{th}}$ is the required SNR threshold.

Using the generalized gain expressions in Section~II and noting that $|f(0,0)|^{2}=1$ under perfect alignment, the boresight directional antenna gains become
\begin{equation}
G_{\mathrm{AP}}(0,0)=N_{1}^{2} G_{0}\approx \frac{4D_1^2G_0}{\lambda^2}, 
\label{eq:aligned_gain_N_AP}
\end{equation}
and
\begin{equation}
    G_{\mathrm{UE}}(0,0)=N_{2}^{2} G_{0}\approx \frac{4D_2^2G_0}{\lambda^2}.
    \label{eq:aligned_gain_N_UE}
\end{equation}

The received SNR at distance $d$ under LoS propagation is therefore
\begin{equation}
\begin{split}
    \mathsf{SNR}(d)&=\frac{P_{\mathrm{t}}\,G_{\mathrm{AP}}(0,0)\,G_{\mathrm{UE}}(0,0)}{N_0}\left( \frac{\lambda}{4\pi d} \right) ^2
\\
&\approx\frac{P_{\mathrm{t}}D_{1}^{2}D_{2}^{2}G_{0}^{2}}{N_0\lambda ^2\pi ^2d^2}
\end{split},
\label{eq:snr_general}
\end{equation}
where $N_{0} = B N_{\mathrm{F}} k T$, with $N_{\mathrm{F}}$ denoting the noise factor, $k$ the Boltzmann constant, and $T$ the system temperature in Kelvin. The parameter $B$ is the signal bandwidth in Hertz.

Evaluating \eqref{eq:snr_general} at $d=d_{\max}$ gives
\begin{equation}
D_1D_2\ge \frac{\lambda \pi d_{\max}}{G_0}\sqrt{\frac{N_{\mathrm{F}}kTB}{P_{\mathrm{t}}}}10^{\frac{\mathsf{SNR}_{\mathrm{th},\mathrm{dB}}}{20}},
\label{eq:cond2_N}
\end{equation}
where $\mathsf{SNR}_{\mathrm{th,dB}}$ is the SNR threshold $\mathsf{SNR}_{\mathrm{th}}$ expressed in decibels (dB).

\textbf{\textit{Contradiction Argument.}}
A far-field THz system must satisfy both Conditions~\eqref{eq:cond1} and~\eqref{eq:cond2_N}, i.e.,
\begin{equation}
\begin{cases}
D_{1} + D_{2} \leq \dfrac{\sqrt{\lambda d_{\min}}}{2}, \\[1.5mm]
D_1D_2\ge \frac{\lambda \pi d_{\max}}{G_0}\sqrt{\frac{N_{\mathrm{F}}kTB}{P_{\mathrm{t}}}}10^{\frac{\mathsf{SNR}_{\mathrm{th},\mathrm{dB}}}{20}}.
\end{cases}
\label{eq:case_constraints}
\end{equation}

To examine whether both conditions can be simultaneously satisfied, we rewrite the second inequality in terms of a single variable. By setting
\begin{equation}
D_{2} = \frac{\sqrt{\lambda d_{\min}}}{2} - D_{1},
\label{eq:D2_subs}
\end{equation}
and substituting~\eqref{eq:D2_subs} into~\eqref{eq:case_constraints}, the second inequality becomes
\begin{equation}
D_{1}\left(\sqrt{\lambda d_{\min}} - 2 D_{1}\right)
\geq 
\frac{2\lambda \pi d_{\max}}{G_0}
\sqrt{Z}
\,10^{\frac{\mathsf{SNR}_{\mathrm{th,dB}}}{20}},
\label{eq:cond2_pre_square_new}
\end{equation}
where $Z$ is set to $Z\triangleq\frac{N_{\mathrm{F}}kTB}{P_{\mathrm{t}}}$ to simplify the notation.

Expanding~\eqref{eq:cond2_pre_square_new}, we obtain a monic quadratic inequality in~$D_{1}$ as
\begin{equation}
D_{1}^{2}
- D_{1}\frac{\sqrt{\lambda d_{\min}}}{2}
+ 
\frac{\lambda \pi d_{\max}}{G_0}
\sqrt{Z}
\,10^{\frac{\mathsf{SNR}_{\mathrm{th,dB}}}{20}}
\;\leq\;0.
\label{eq:cond2_quadratic}
\end{equation}

Let $p = -{\sqrt{\lambda d_{\min}}}/{2}$ and $q =
\lambda \pi d_{\max}
\sqrt{Z}
\,10^{\frac{\mathsf{SNR}_{\mathrm{th,dB}}}{20}}/G_0,$
so that~\eqref{eq:cond2_quadratic} takes the standard form
\begin{equation}
    D_{1}^{2} + pD_{1} + q \leq 0.
    \label{eq:cond2_simplify}
\end{equation}

This has up to two real roots, denoted by $x_{1}$ and $x_{2}$ with $x_{1} \le x_{2}$. Since the left-hand side must be non-positive, the feasible values of $D_{1}$ necessarily satisfy $D_{1} \in [x_{1}, x_{2}].$ Because $D_{1}$ represents a physical aperture size, the additional constraint $D_{1}>0$ must always hold. Therefore, the sign of $x_{1}$ becomes relevant in determining whether a physically meaningful solution exists. To characterize $x_{1}$ and $x_{2}$, we employ Vieta’s formulas~\cite{herstein1996abstract}. In particular, the roots satisfy
\begin{equation}
\begin{cases}
x_{1} + x_{2} = -p = \dfrac{\sqrt{\lambda d_{\min}}}{2},\\[2mm]
x_{1} x_{2} = q = 
\frac{\lambda \pi d_{\max}}{G_0}
\sqrt{\dfrac{N_{\mathrm{F}} k T B}{P_{\mathrm{t}}}}
\, 10^{\frac{\mathsf{SNR}_{\mathrm{th,dB}}}{20}} .
\end{cases}
\label{eq:vieta_new}
\end{equation}

From the right-hand sides of~\eqref{eq:vieta_new}, we observe that both $x_{1}$ and $x_{2}$ (if real) must be non-negative:  
(1) the sum $x_{1}+x_{2}$ is non-negative since it equals $\sqrt{\lambda d_{\min}}/2$;  
(2) the product $x_{1} x_{2} = q$ is strictly positive because all terms in~$q$ are positive.  
Hence, if real roots exist, they satisfy $0 \le x_{1} \le x_{2},$ which guarantees that the feasible interval $D_{1} \in [x_{1}, x_{2}]$ includes physically meaningful aperture sizes with $D_{1} > 0$.

\textit{Consequently, the existence of any far-field–compatible aperture pair $(D_{1},D_{2})$ now depends only on whether the quadratic function admits real roots, i.e., whether its discriminant is non-negative.} Using the standard discriminant condition,
\begin{equation}
    \left(\frac{p}{2}\right)^{2} - q \ge 0,
\end{equation}
we obtain the necessary and sufficient feasibility requirement
\begin{equation}
\left(\frac{\sqrt{\lambda d_{\min}}}{4}\right)^{2}
-
\frac{\lambda \pi d_{\max}}{G_0}
\sqrt{\frac{N_{\mathrm{F}}kTB}{P_{\mathrm{t}}}}
\,10^{\frac{\mathsf{SNR}_{\mathrm{th,dB}}}{20}}
\geq0.
\label{eq:discriminant_new}
\end{equation}

Rearranging~\eqref{eq:discriminant_new} yields a constraint on the usable bandwidth $B$, which is given by
\begin{equation}
B\le \frac{P_{\mathrm{t}}G_{0}^{2}d_{\min}^{2}}{256\pi ^2d_{\max}^{2}N_{\mathrm{F}}kT}10^{-\frac{\mathsf{SNR}_{\mathrm{th},\mathrm{dB}}}{10}}.
\label{eq:bw_new}
\end{equation}

The equality case of~\eqref{eq:bw_new} corresponds to the maximal feasible bandwidth and occurs when the quadratic has exactly one root, i.e., $D_{1} = x_{1} = x_{2}$. From Vieta’s formula,
\begin{equation}
    x_{1}+x_{2}=\frac{\sqrt{\lambda d_{\min}}}{2},
\end{equation}
we obtain the aperture values
\begin{equation}
D_{1}=D_{2}=\frac{\sqrt{\lambda d_{\min}}}{4}.
\label{eq:D_opt_new}
\end{equation}

This symmetric configuration is, however, not representative of mobile THz systems, where typically $D_{2}\!\ll\!D_{1}$. It instead corresponds to stationary backhaul-type links with a fixed distance $d_{\min}=d_{\max}=d$. For such stationary links, the maximal feasible bandwidth is obtained by substituting~\eqref{eq:D_opt_new} into~\eqref{eq:cond2_N}, given as
\begin{equation}
B_{\mathrm{Stationary}}^{(\max)}\le \frac{G_{0}^{2}}{256\pi ^2kT}10^{\frac{P_{\mathrm{t},\mathrm{dBm}}-\mathsf{SNR}_{\mathrm{th},\mathrm{dB}}-N_{\mathrm{F,dB}}-30}{10}},
\label{eq:bw_final_stationary_new}
\end{equation}
where $P_{\mathrm{t,dBm}}$ is the transmit power in dBm, and $N_{\mathrm{F,dB}}$ is the receiver noise figure in dB.

\textit{Thus, a \textbf{stationary} THz link may operate in the far field only if its system bandwidth does not exceed the limit in~\eqref{eq:bw_final_stationary_new}.}

Equivalently, by defining the transmit power spectral density (PSD) as $S_{\mathrm{t}} \triangleq {P_t}/{B}$ (or $S_{\mathrm{t,dBm/Hz}}\triangleq P_{\mathrm{t,dBm}}-10\log_{10}(B)$) the same condition can be rewritten as a minimum required PSD, i.e.,
\begin{equation}
\begin{split}
    S_{\mathrm{t,dBm/Hz}}
&\ge
64
+\mathsf{SNR}_{\mathrm{th,dB}}
+N_{\mathrm{F,dB}}+10\log_{10}(kT)\\
&-20\log_{10}(G_0).
\end{split}
\label{eq:tx_psd_cond_dbm}
\end{equation}

We now turn to the \textit{\textbf{mobile case}}, where the communication distance varies within $[d_{\min}, d_{\max}]$ and the AP and UE employ apertures of different sizes. 

By substituting $D_{1}=L D_{2}$ and $d_{\max}=Md_{\min}$ into the far-field and SNR constraints in~\eqref{eq:case_constraints}, the joint feasibility conditions become
\begin{subnumcases}{\label{eq:mobile_joint_constraints}}
(L+1)D_{2} \le \dfrac{\sqrt{\lambda d_{\min}}}{2}, \label{eq:constraint_a} \\[1.5mm]
L D_{2}^{2} \ge \dfrac{\lambda \pi M d_{\min}}{G_{0}} \sqrt{\dfrac{N_{\mathrm{F}} k T B}{P_{\mathrm{t}}}} \,10^{\frac{\mathsf{SNR}_{\mathrm{th,dB}}}{20}}. \label{eq:constraint_b}
\end{subnumcases}

From \eqref{eq:constraint_a}, the UE aperture is upper bounded as
\begin{equation}
D_{2} \le \frac{\sqrt{\lambda d_{\min}}}{2(L+1)}.
\label{eq:D2_upper}
\end{equation}

Substituting~\eqref{eq:D2_upper} into \eqref{eq:constraint_b} and solving for~$B$ yields the maximal feasible bandwidth for a mobile THz system as
\begin{equation}
B_{\mathrm{Mobile}}
\;\le\;
\frac{G_0^2L^{2}}
{16 \pi^2k TM^{2} (L+1)^{4}}
\,
10^{\frac{P_{\mathrm{t,dBm}}
-\mathsf{SNR}_{\mathrm{th,dB}}
-N_{\mathrm{F,dB}}
-30}{10}}.
\label{eq:bw_mobile_general}
\end{equation}

A comparison with the stationary limit in~\eqref{eq:bw_final_stationary_new} shows that mobility introduces two multiplicative penalties:  
(i) a \textit{mobility penalty} of $M^{2}$ arising from the variation in the communication distance, and  
(ii) an \textit{aperture-imbalance penalty} of $(L+1)^{4}/(16L^{2})$ that reflects the asymmetry between the AP and UE apertures. Notably, when $M = 1$ and $L = 1$ (i.e., when the link distance is fixed and the AP and UE employ equal-size apertures), the two penalties disappear, and~\eqref{eq:bw_mobile_general} reduces exactly to the stationary limit in~\eqref{eq:bw_final_stationary_new}.

By expressing~\eqref{eq:bw_mobile_general} relative to the stationary limit \eqref{eq:bw_final_stationary_new}, we obtain
\begin{equation}
B_{\mathrm{Mobile}}^{(\max)}
={
B_{\mathrm{Stationary}}^{(\max)}
}/{
\left(M^{2}\,\dfrac{(L+1)^{4}}{16L^{2}}\right)
},
\label{eq:bw_mobile_ratio}
\end{equation}
which provides an explicit scaling law describing how mobility and antenna asymmetry reduce the far-field feasibility region.

\textit{Thus, a mobile THz system can operate in the far field only when its bandwidth is below the threshold in~\eqref{eq:bw_mobile_general} (or equivalently~\eqref{eq:bw_mobile_ratio}), which highlights the combined impact of mobility and aperture asymmetry on far-field feasibility.}

By inverting \eqref{eq:bw_mobile_general}, the minimum transmit power required to support a target bandwidth $B$ while operating \emph{exclusively} in the far field is
\begin{equation}
\begin{split}
&P_{\mathrm{t},\mathrm{dBm}}=52+\mathsf{SNR}_{\mathrm{th},\mathrm{dB}}+N_{\mathrm{F},\mathrm{dB}}+10\log _{10}\left( kT \right) \\&+10\log _{10}\left( B_{\mathrm{Mobile}}^{\left( \max \right)} \right) +20\log _{10}\left( \frac{M(L+1)^2}{G_0L} \right). 
\end{split}
\label{eq:Pt_1}
\end{equation}

\subsection{Scenario 1: UE Rotation With One Angle}
\label{sec:scenario1}

\begin{figure}[!t]
    \centering
    \includegraphics[width=0.95\linewidth]{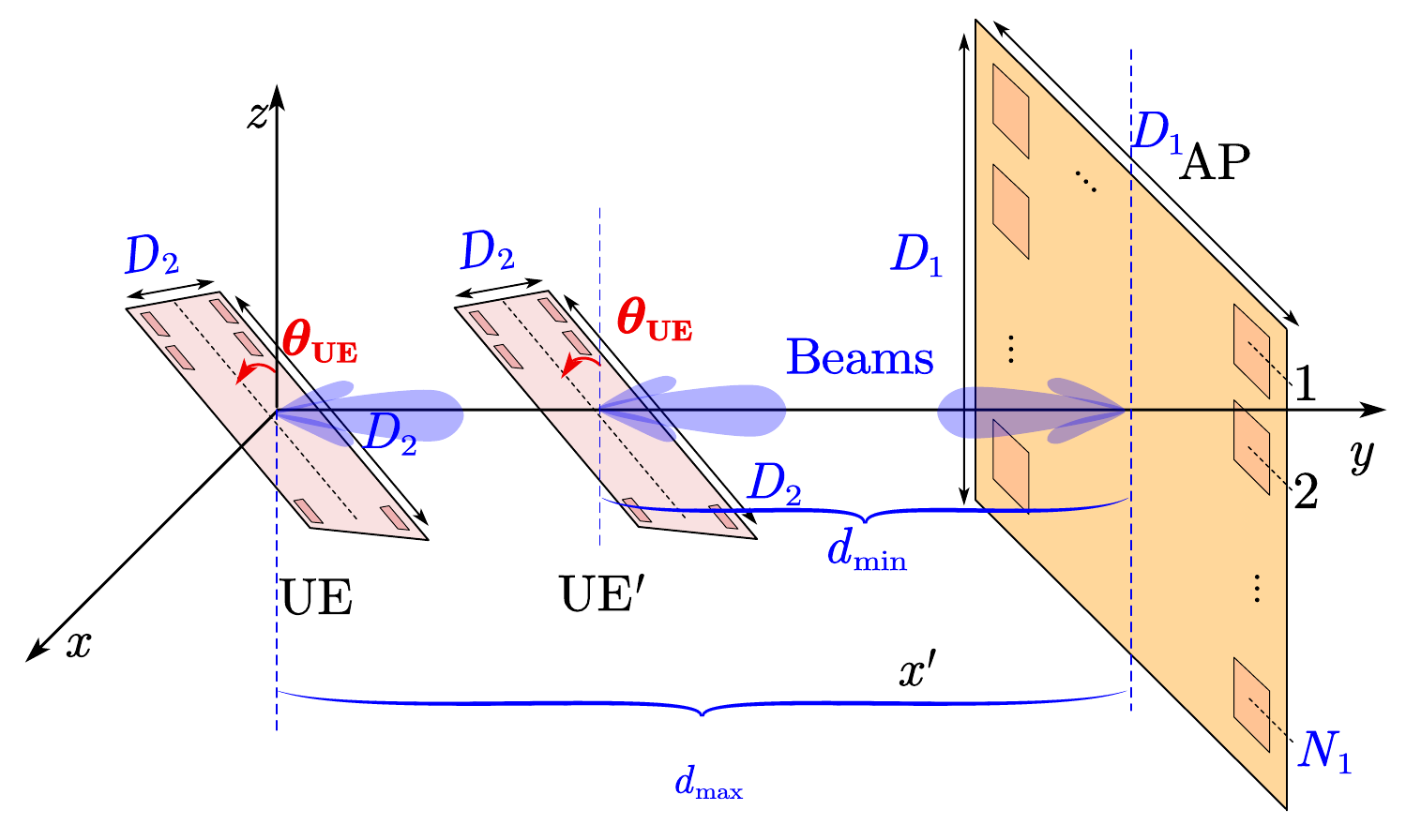}
    \caption{Scenario~1: UE rotated by a single angle $\theta_{\mathrm{UE}}$ around the $x$-axis.}
    \label{fig:scenario1}
\end{figure}

We now consider \textit{Scenario~1}, illustrated in Fig.~\ref{fig:scenario1}, where the UE is rotated by a single angle $\theta_{\mathrm{UE}}$ around the $x$-axis. Throughout Scenarios~1--4, we assume that the rotated array applies \emph{pre-steering} so that its main lobe is always directed toward the counterpart node along the LoS despite the physical rotation~\cite{zhang2025impact}. Therefore, the impact of rotation is captured through (i) the modified near-field boundary distance and (ii) the element-pattern attenuation term $|f(\cdot,\cdot)|^2$.

We first recall the near-field distance for UPA–UPA case with one UE rotation angle $\theta_{\mathrm{UE}}$, as derived in~\cite{zhang2025misalignedTWC} as
\begin{equation}\label{equ:UPA1_near_field}
    d_{\mathrm{F}}(\theta_{\mathrm{UE}})
    =
    \frac{
        2\left[ (D_{1}+D_{2})^{2} + (D_{1}+D_{2}\cos\theta_{\mathrm{UE}})^{2} \right]
    }{\lambda}.
\end{equation}

Comparing~\eqref{equ:UPA1_near_field} with the aligned case in~\eqref{eq:dF_general} shows that rotation \textit{reduces} the near-field boundary distance. As a result, \textit{Condition~1}, $d_{\min} \ge d_{\mathrm{F}}(\theta_{\mathrm{UE}}),$
becomes \textit{easier} to satisfy in the presence of UE rotation. However, rotation also modifies the array gain through the element radiation pattern. Using the generalized expression in Section~II, the directional antenna gain of UE becomes
\begin{equation}
G_{\mathrm{UE}}(\theta_{\mathrm{UE}}, 0)
=
N_{2}^{2} G_{0}\, |f(\theta_{\mathrm{UE}}, 0)|^{2}.
\label{eq:G_UE_rot}
\end{equation}

Since the element pattern satisfies $|f(\theta,\phi)|^2 \le |f(0,0)|^2$, any rotation introduces a gain loss relative to the aligned case. This loss has opposite effect on \textit{Condition~2}, $\mathsf{SNR}(d_{\max}) \ge \mathsf{SNR}_{\mathrm{th}},$ which becomes more \textit{difficult} to satisfy.

Therefore, UE rotation simultaneously (i) relaxes \textit{Condition~1} by reducing the near-field threshold distance, but (ii) tightens \textit{Condition~2} by degrading the effective array gain. Both effects must be jointly accounted for to determine the maximal bandwidth under mobility.

Substituting $D_{1}=L D_{2}$ into~\eqref{equ:UPA1_near_field} and imposing \textit{Condition~1}, $d_{\min} \ge d_{\mathrm{F}}(\theta_{\mathrm{UE}})$, we have
\begin{equation}
\big((L+1)^{2} + (L+\cos\theta_{\mathrm{UE}})^{2}\big) D_{2}^{2}
\le
\frac{\lambda d_{\min}}{2}.
\label{eq:cond1_rot_LM}
\end{equation}

For \textit{Condition~2}, we substitute $G_{\mathrm{AP}}(0,0)=N_{1}^{2}G_{0}$ and $G_{\mathrm{UE}}(\theta_{\mathrm{UE}},0)$ from~\eqref{eq:G_UE_rot} into the SNR expression and evaluate it at $d=d_{\max}$, which gives
\begin{equation}
\mathsf{SNR}(d_{\max})
=
\frac{P_{\mathrm{t}} N_{1}^{2} N_{2}^{2} G_{0}^{2} |f(\theta_{\mathrm{UE}},0)|^{2}}{N_{0}}
\left(\frac{\lambda}{4\pi d_{\max}}\right)^{2}.
\end{equation}

Imposing $\mathsf{SNR}(d_{\max}) \ge \mathsf{SNR}_{\mathrm{th}}$, the above condition is equivalent to
\begin{equation}
D_{1} D_{2}
\ge
\frac{\lambda \pi d_{\max}}{G_{0} |f(\theta_{\mathrm{UE}},0)|}
\sqrt{\frac{N_{\mathrm{F}} k T B}{P_{\mathrm{t}}}}
\,10^{\frac{\mathsf{SNR}_{\mathrm{th,dB}}}{20}}.
\label{eq:cond2_rot_D1D2}
\end{equation}

We now express \eqref{eq:cond2_rot_D1D2} in terms of $L$ and $M$ by substituting $D_{1}=L D_{2}$ and $d_{\max}=M d_{\min}$ as
\begin{equation}
L D_{2}^{2}
\ge
\frac{\lambda \pi M d_{\min}}{G_{0} |f(\theta_{\mathrm{UE}},0)|}
\sqrt{\frac{N_{\mathrm{F}} k T B}{P_{\mathrm{t}}}}
\,10^{\frac{\mathsf{SNR}_{\mathrm{th,dB}}}{20}}.
\label{eq:cond2_rot_LM}
\end{equation}

Combining \eqref{eq:cond1_rot_LM} and \eqref{eq:cond2_rot_LM}, and eliminating $D_{2}^{2}$ in the same way as in Scenario~0, we obtain an upper bound on the feasible bandwidth as
\begin{equation}
\begin{split}
B_{\mathrm{Mobile,UE}}^{(\max)}(\theta_{\mathrm{UE}})
\le&
\frac{L^{2} G_{0}^{2} |f(\theta_{\mathrm{UE}},0)|^{2}}{
4\pi^{2} k T M^{2}
\big((L+1)^{2} + (L+\cos\theta_{\mathrm{UE}})^{2}\big)^{2}
}\\
\times&10^{\frac{P_{\mathrm{t,dBm}} - \mathsf{SNR}_{\mathrm{th,dB}} - N_{\mathrm{F,dB}} - 30}{10}}.
\end{split}
\label{eq:bw_mobile_rot_final}
\end{equation}

When $\theta_{\mathrm{UE}}=0$ and $|f(0,0)|^{2}=1$, the geometric factor in the denominator satisfies $\big((L+1)^{2} + (L+\cos 0)^{2}\big)^{2} = 4(L+1)^{4},$ and \eqref{eq:bw_mobile_rot_final} reduces exactly to the aligned mobile bound in~\eqref{eq:bw_mobile_general}. 

By expressing the maximum bandwidth in Scenario~1 relative to the stationary limit~\eqref{eq:bw_final_stationary_new}, we obtain
\begin{equation}
B_{\mathrm{Mobile,UE}}^{(\max)}(\theta_{\mathrm{UE}})
=
\frac
{B_{\mathrm{Stationary}}^{(\max)}
}{
\Psi_{1}(\theta_{\mathrm{UE}};L,M)
},
\label{eq:bw_rot_ratio_template}
\end{equation}
where 
\begin{equation}
    \Psi_{1}(\theta_{\mathrm{UE}};L,M)= \frac{M^2\bigl( (L+1)^2+(L+\cos \theta _{\mathrm{UE}})^2 \bigr) ^2}{64L^2|f(\theta _{\mathrm{UE}},0)|^2}
\end{equation}
denotes the multiplicative penalty induced jointly by (i) the geometric rotation of the UE and (ii) the element-pattern loss. Notably, in the special case $\theta_{\mathrm{UE}}=0$, we have $|f(0,0)|^{2}=1$ and $\Psi_{1}(0;L,M)=M^{2}(L+1)^{4}/(16L^{2})$, which exactly recovers the aligned-aperture mobile limit in~\eqref{eq:bw_mobile_ratio}.

By inverting~\eqref{eq:bw_mobile_rot_final} and substituting~\eqref{eq:bw_rot_ratio_template}, the minimum transmit power required to support a target bandwidth $B$ while operating \emph{exclusively} in the far field is in~\eqref{equ:Pt2} on the top of the next page.
\begin{figure*}[!t]
\begin{equation}\label{equ:Pt2}
    P_{\mathrm{t},\mathrm{dBm}}^{\mathrm{UE}1}\left( \theta _{\mathrm{UE}} \right) =64+\mathsf{SNR}_{\mathrm{th},\mathrm{dB}}+N_{\mathrm{F},\mathrm{dB}}+10\log _{10}\left( kT \right) +10\log _{10}\left( B_{\mathrm{Mobile},\mathrm{UE}}^{(\max\mathrm{)}}(\theta _{\mathrm{UE}}) \right) +10\log _{10}\left( \Psi _1\left( \theta _{\mathrm{UE}};L,M \right) \right).
\end{equation}
\hrulefill
\end{figure*}

\subsection{Scenario 2: UE Rotation With Two Angles}
\label{sec:scenario2}

\begin{figure}[!t]
    \centering
    \includegraphics[width=0.95\linewidth]{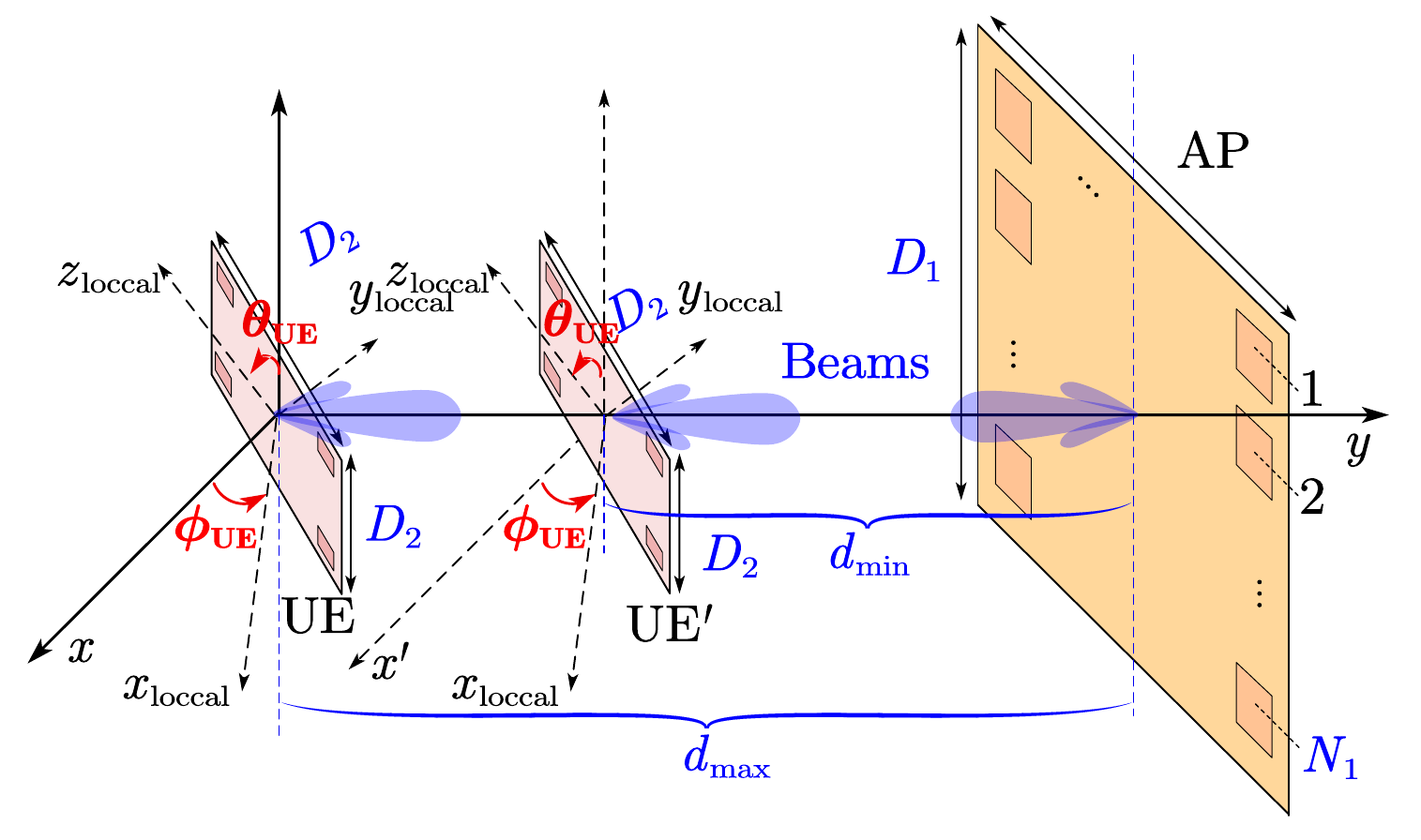}
    \caption{Scenario~2: UE rotated by two angles, $\theta_{\mathrm{UE}}$ around the $x$-axis and $\phi_{\mathrm{UE}}$ around the $z$-axis.}
    \label{fig:scenario2}
\end{figure}

We now extend Scenario~1 by allowing the UE to rotate around two axes, as illustrated in Fig.~\ref{fig:scenario2}. Specifically, the UE undergoes a composite rotation $\mathbf{R}(\theta_{\mathrm{UE}},\phi_{\mathrm{UE}})=\mathbf{R}_{z}(\phi_{\mathrm{UE}})\mathbf{R}_{x}(\theta_{\mathrm{UE}})$, where the component rotation matrices are
\begin{equation}
\mathbf{R}_{x}(\theta_{\mathrm{UE}})
=
\begin{bmatrix}
1 & 0 & 0 \\
0 & \cos\theta_{\mathrm{UE}} & -\sin\theta_{\mathrm{UE}} \\
0 & \sin\theta_{\mathrm{UE}} & \cos\theta_{\mathrm{UE}}
\end{bmatrix},
\end{equation}
\begin{equation}
    \mathbf{R}_{z}(\phi_{\mathrm{UE}})
=
\begin{bmatrix}
\cos\phi_{\mathrm{UE}} & -\sin\phi_{\mathrm{UE}} & 0 \\
\sin\phi_{\mathrm{UE}} & \cos\phi_{\mathrm{UE}} & 0 \\
0 & 0 & 1
\end{bmatrix}.
\end{equation}

For the UPA–UPA case with two UE rotation angles $(\theta_{\mathrm{UE}},\phi_{\mathrm{UE}})$, the near-field distance can be written as~\cite{zhang2025impact,zhang2025misalignedTWC}
\begin{equation}
\begin{split}
&d_{\mathrm{F}}(\theta _{\mathrm{UE}},\phi _{\mathrm{UE}})=\frac{2\left( D_1+D_2\cos \theta _{\mathrm{UE}} \right) ^2}{\lambda}\\&+\frac{2\left( D_1+D_2\left( \cos \phi _{\mathrm{UE}}+|\sin \theta _{\mathrm{UE}}\sin \phi _{\mathrm{UE}}| \right) \right) ^2}{\lambda}.
\end{split}
\label{eq:dF_two_angles}
\end{equation}

Imposing \textit{Condition~1}, $d_{\min} \ge d_{\mathrm{F}}(\theta_{\mathrm{UE}},\phi_{\mathrm{UE}})$, and substituting $D_{1} = L D_{2}$ with $L = D_{1}/D_{2}$ yield
\begin{equation}
F(L,\theta_{\mathrm{UE}},\phi_{\mathrm{UE}})\,D_{2}^{2}
\le
\frac{ \lambda d_{\min}}{2},
\label{eq:cond1_two_LM}
\end{equation}
where
\begin{equation}
\begin{split}
F(L,\theta_{\mathrm{UE}},\phi_{\mathrm{UE}})
=
\big(L + \cos\theta_{\mathrm{UE}}\big)^{2}
+\\
\big(L + \cos\phi_{\mathrm{UE}} + |\sin\theta_{\mathrm{UE}}\sin\phi_{\mathrm{UE}}|\big)^{2}.
\end{split}
\label{eq:F_two_angles_def}
\end{equation}

On the array-gain side, the AP remains aligned with gain $G_{\mathrm{AP}}(0,0) = N_{1}^{2}G_{0}$, while the UE gain becomes
\begin{equation}
G_{\mathrm{UE}}(\theta_{\mathrm{UE}},\phi_{\mathrm{UE}})=N_{2}^{2} G_{0}\, |f(\theta_{\mathrm{UE}},\phi_{\mathrm{UE}})|^{2}.
\label{eq:G_UE_two_angles}
\end{equation}

Substituting \eqref{eq:G_UE_two_angles} and $G_{\mathrm{AP}}(0,0)$ into the SNR expression and evaluating it at $d=d_{\max}$, we impose \textit{Condition~2}, $\mathsf{SNR}(d_{\max}) \ge \mathsf{SNR}_{\mathrm{th}}$, and obtain after rearrangement
\begin{equation}
D_{1}D_{2}
\ge
\frac{\lambda \pi d_{\max}}{G_{0} |f(\theta_{\mathrm{UE}},\phi_{\mathrm{UE}})|}
\sqrt{\frac{N_{\mathrm{F}} k T B}{P_{\mathrm{t}}}}
\,10^{\frac{\mathsf{SNR}_{\mathrm{th,dB}}}{20}}.
\label{eq:cond2_two_D1D2}
\end{equation}
Using $D_{1} = L D_{2}$ and $d_{\max} = M d_{\min}$ with $M = d_{\max}/d_{\min}$, \eqref{eq:cond2_two_D1D2} becomes
\begin{equation}
L D_{2}^{2}
\ge
\frac{\lambda \pi M d_{\min}}{G_{0} |f(\theta_{\mathrm{UE}},\phi_{\mathrm{UE}})|}
\sqrt{\frac{N_{\mathrm{F}} k T B}{P_{\mathrm{t}}}}
\,10^{\frac{\mathsf{SNR}_{\mathrm{th,dB}}}{20}}.
\label{eq:cond2_two_LM}
\end{equation}

Combining \eqref{eq:cond1_two_LM} and \eqref{eq:cond2_two_LM} and eliminating $D_{2}^{2}$ in the same manner as in Scenarios~0 and~1, we obtain an upper bound on the bandwidth of a mobile THz system with two UE rotation angles as
\begin{equation}
\begin{split}
B_{\mathrm{Mobile},\mathrm{UE}2}^{(\max\mathrm{)}}(\theta _{\mathrm{UE}},\phi _{\mathrm{UE}})&\le \frac{L^2G_{0}^{2}|f(\theta _{\mathrm{UE}},\phi _{\mathrm{UE}})|^2}{4\pi ^2kTM^2F^2(L,\theta _{\mathrm{UE}},\phi _{\mathrm{UE}})}\,
\\
&\times 10^{\frac{P_{\mathrm{t},\mathrm{dBm}}-\mathsf{SNR}_{\mathrm{th},\mathrm{dB}}-N_{\mathrm{F},\mathrm{dB}}-30}{10}}.
\end{split}
\label{eq:bw_mobile_two_angles}
\end{equation}

Expression~\eqref{eq:bw_mobile_two_angles} generalizes the one-angle result in~\eqref{eq:bw_mobile_rot_final} by incorporating both the composite-rotation geometry through $F(L,\theta_{\mathrm{UE}},\phi_{\mathrm{UE}})$ and the two-dimensional element-pattern attenuation $|f(\theta_{\mathrm{UE}},\phi_{\mathrm{UE}})|^{2}$. When $\phi_{\mathrm{UE}} = 0$, \eqref{eq:bw_mobile_two_angles} reduces to the single-angle case in Scenario~1 , while setting $\theta_{\mathrm{UE}} = \phi_{\mathrm{UE}} = 0$ and $|f(0,0)|^{2}=1$ recovers the aligned mobile bound in~\eqref{eq:bw_mobile_general}.

Similarly to Scenario~1, expressing~\eqref{eq:bw_mobile_two_angles} relative to the stationary limit~\eqref{eq:bw_final_stationary_new}, we obtain
\begin{equation}
B_{\mathrm{Mobile,UE2}}^{(\max)}(\theta_{\mathrm{UE}},\phi_{\mathrm{UE}})
=
\frac{
B_{\mathrm{Stationary}}^{(\max)}
}{
\Psi_{2}(\theta_{\mathrm{UE}},\phi_{\mathrm{UE}};L,M)
},
\label{eq:bw_rot2_ratio}
\end{equation}
where
\begin{equation}
\Psi_{2}(\theta_{\mathrm{UE}},\phi_{\mathrm{UE}};L,M)
=
\frac{
M^{2}\, F^{2}(L,\theta_{\mathrm{UE}},\phi_{\mathrm{UE}})
}{
64\,L^{2}\, |f(\theta_{\mathrm{UE}},\phi_{\mathrm{UE}})|^{2}
},
\label{eq:Psi2_def}
\end{equation}
characterizes the multiplicative degradation due to simultaneous UE rotations in two angular dimensions. The function $F(L,\theta_{\mathrm{UE}},\phi_{\mathrm{UE}})$ captures the geometric distortion of the near-field boundary, whereas the factor $|f(\theta_{\mathrm{UE}},\phi_{\mathrm{UE}})|^{2}$ accounts for the corresponding element-pattern loss.

As expected, when $(\theta_{\mathrm{UE}},\phi_{\mathrm{UE}})=(0,0)$, we have $F(L,0,0)=2{(L+1)^{2}}$ and $|f(0,0)|^{2}=1$, which yields $\Psi_{2}(0,0;L,M)=M^{2}(L+1)^{4}/(16L^{2})$, thereby reducing Scenario~2 to the aligned-aperture mobile limit in~\eqref{eq:bw_mobile_ratio}.

By inverting~\eqref{eq:bw_mobile_two_angles} and substituting~\eqref{eq:Psi2_def}, the minimum transmit power required to support a target bandwidth $B$ while operating \emph{exclusively} in the far field is 
\begin{equation}
    \begin{split}
        &P_{\mathrm{t},\mathrm{dBm}}^{\mathrm{UE}2}\left( \theta _{\mathrm{UE}},\phi _{\mathrm{UE}} \right) =64+\mathsf{SNR}_{\mathrm{th},\mathrm{dB}}+N_{\mathrm{F},\mathrm{dB}}\\&+10\log _{10}\left( kT \right) +10\log _{10}\left( B_{\mathrm{Mobile},\mathrm{UE}}^{(\max)}(\theta _{\mathrm{UE}},\phi _{\mathrm{UE}}) \right) \\&+10\log _{10}\left( \Psi _2\left( \theta _{\mathrm{UE}},\phi _{\mathrm{UE}};L,M \right) \right).
    \end{split}
\end{equation}

\subsection{Scenario 3: Relative Angular Variation With One Angle}
\label{sec:scenario3}

\begin{figure}[!t]
    \centering
    \includegraphics[width=0.95\linewidth]{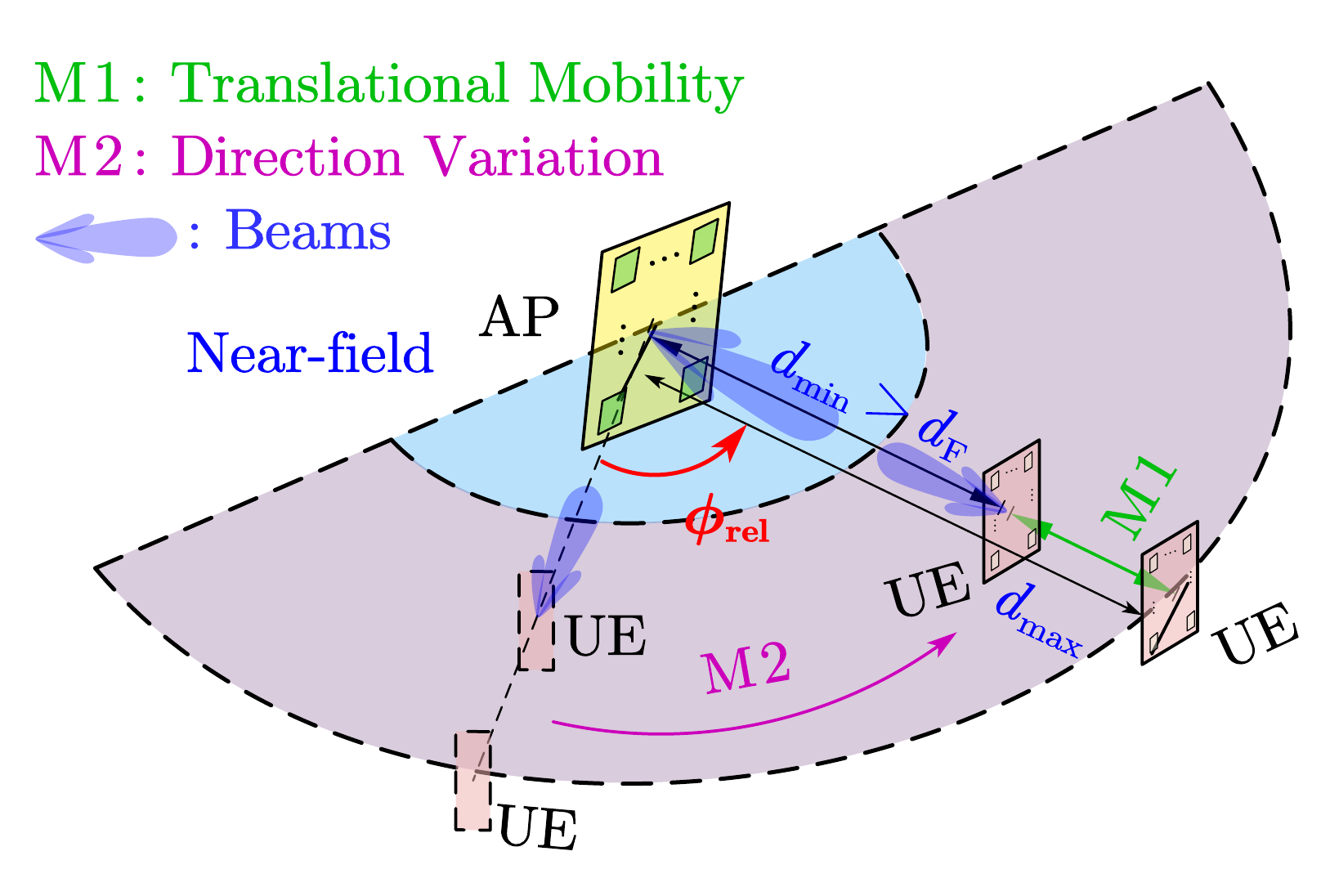}
    \caption{Scenario~3: The UE maintains its array orthogonal to the instantaneous LoS direction and UE motion induces a one-angle \emph{relative} angular variation $\phi_{\mathrm{rel}}$ in the horizontal plane.}
    \label{fig:scenario3}
\end{figure}

Having characterized UE \emph{rotation} in Scenarios~1 and~2, we now consider \emph{relative angular variation} induced by UE motion. 
In \textit{Scenario~3}, illustrated in Fig.~\ref{fig:scenario3}, the UE does not self-rotate, instead, the UE is assumed to maintain its array orthogonal to the instantaneous LoS direction. As the UE moves within the coverage region, the LoS direction (and hence the relative pointing direction with respect to the fixed AP boresight) varies, which can be parameterized by a single relative angle $\phi_{\mathrm{rel}}$ in the horizontal plane.

For the UPA--UPA case with one AP rotation angle $\phi_{\mathrm{rel}}$, the near-field distance can be written as~\cite{zhang2025misalignedTWC}
\begin{equation}
d_{\mathrm{F}}^{(\mathrm{AP})}(\phi_{\mathrm{rel}})
=
\frac{2 (D_{1}+D_{2})^{2}}{\lambda}
+
\frac{2 (D_{1}\cos\phi_{\mathrm{rel}} + D_{2})^{2}}{\lambda}.
\label{eq:dF_AP}
\end{equation}

As before, imposing the far-field requirement $d_{\min} \ge d_{\mathrm{F}}^{(\mathrm{AP})}(\phi_{\mathrm{rel}})$ and substituting $D_{1} = L D_{2}$ yield
\begin{equation}
\big[(L+1)^{2} + (L\cos\phi_{\mathrm{rel}}+1)^{2}\big] D_{2}^{2}
\le
\frac{\lambda d_{\min}}{2}.
\label{eq:cond1_AP_L}
\end{equation}

Regarding the array gain, the UE remains aligned and thus retains $G_{\mathrm{UE}}(0,0) = N_{2}^{2} G_{0},$ whereas the AP experiences the rotated gain
\begin{equation}
G_{\mathrm{AP}}(0,\phi_{\mathrm{rel}})=N_{1}^{2} G_{0}\, |f(0,\phi_{\mathrm{rel}})|^{2},
\label{eq:G_AP_rot}
\end{equation}
due to the element pattern. Substituting \eqref{eq:G_AP_rot} and $G_{\mathrm{UE}}(0,0)$ into the SNR expression and enforcing \textit{Condition~2}, $\mathsf{SNR}(d_{\max}) \ge \mathsf{SNR}_{\mathrm{th}}$, we have
\begin{equation}
D_{1} D_{2}
\ge
\frac{\lambda \pi d_{\max}}{G_{0} |f(0,\phi_{\mathrm{rel}})|}
\sqrt{\frac{N_{\mathrm{F}} k T B}{P_{\mathrm{t}}}}
\,10^{\frac{\mathsf{SNR}_{\mathrm{th,dB}}}{20}}.
\label{eq:cond2_AP}
\end{equation}

Using $D_{1} = L D_{2}$ and $d_{\max} = M d_{\min}$ yields
\begin{equation}
L D_{2}^{2}
\ge
\frac{\lambda \pi M d_{\min}}{G_{0} |f(0,\phi_{\mathrm{rel}})|}
\sqrt{\frac{N_{\mathrm{F}} k T B}{P_{\mathrm{t}}}}
\,10^{\frac{\mathsf{SNR}_{\mathrm{th,dB}}}{20}}.
\label{eq:cond2_AP_L}
\end{equation}

Combining \eqref{eq:cond1_AP_L} and \eqref{eq:cond2_AP_L} and eliminating $D_{2}^{2}$ in the same manner as in Scenarios~0–2, we obtain the maximum feasible bandwidth under AP-side rotation
\begin{equation}
\begin{split}
    B_{\mathrm{Mobile,AP}}^{(\max)}(\phi_{\mathrm{rel}})
&\le
\frac{
L^{2} G_{0}^{2} |f(0,\phi_{\mathrm{rel}})|^{2}
}{
4\pi^{2} k T M^{2} \big((L+1)^{2} + (L\cos\phi_{\mathrm{rel}}+1)^{2}\big)^{2}
}
\\
&\times10^{\frac{P_{\mathrm{t,dBm}} - \mathsf{SNR}_{\mathrm{th,dB}} - N_{\mathrm{F,dB}} - 30}{10}}.
\end{split}
\label{eq:bw_AP_final}
\end{equation}

\begin{figure}[!t]
    \centering
    \includegraphics[width=0.9\linewidth]{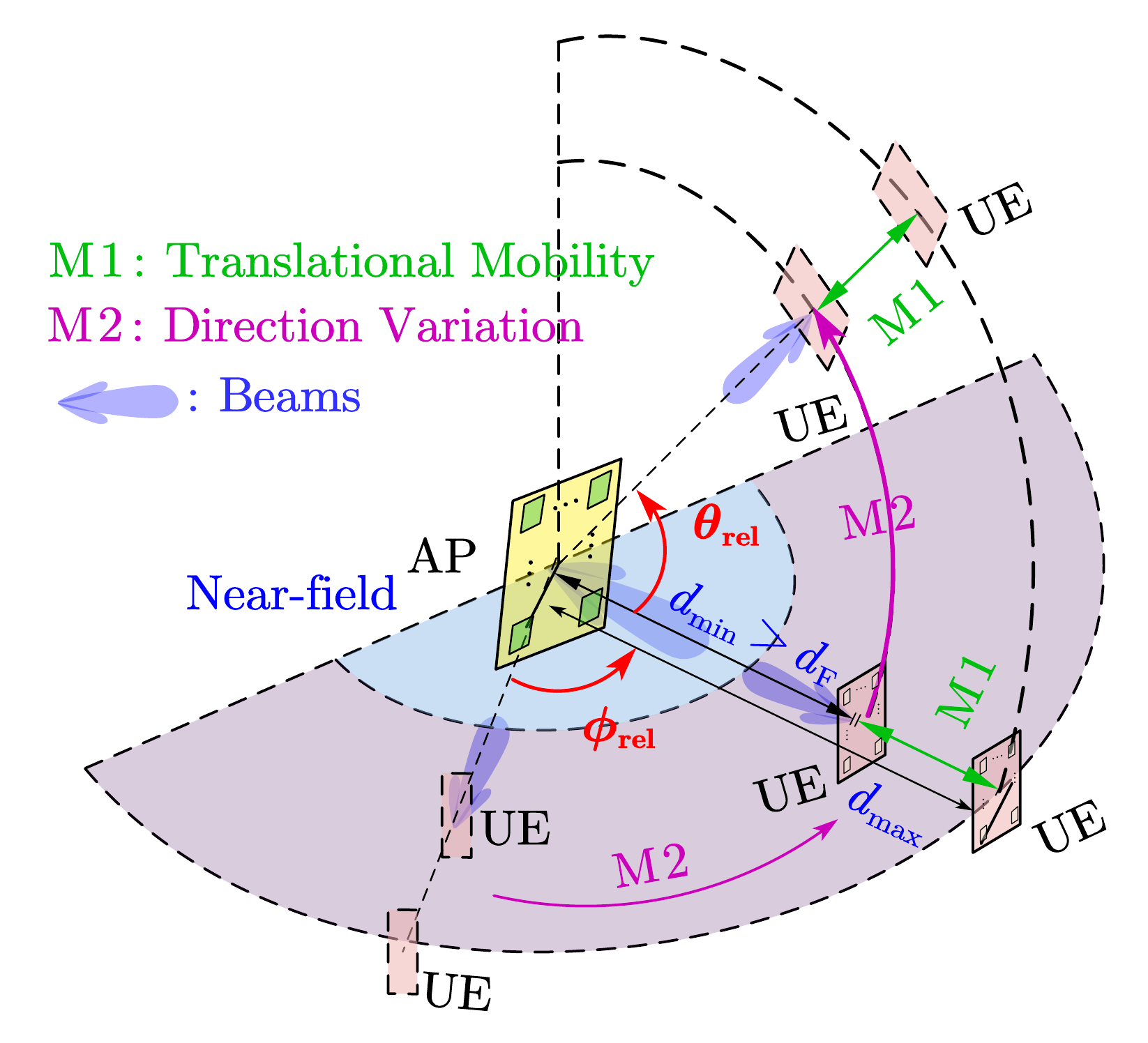}
    \caption{Scenario~4: The UE maintains its array orthogonal to the instantaneous LoS direction and UE motion induces a two-angle \textit{direction variation} $(\theta_{\mathrm{rel}},\phi_{\mathrm{rel}})$.}
    \label{fig:scenario4}
\end{figure}

By expressing the maximum bandwidth \eqref{eq:bw_AP_final} in Scenario~3 relative to the stationary limit~\eqref{eq:bw_final_stationary_new}, we obtain
\begin{equation}
B_{\mathrm{Mobile,AP}}^{(\max)}(\phi_{\mathrm{rel}})
=
\frac{
B_{\mathrm{Stationary}}^{(\max)}
}{
\Psi_{3}(\phi_{\mathrm{rel}};L,M)
},
\label{eq:bw_AP_ratio}
\end{equation}
where
\begin{equation}
\Psi_{3}(\phi_{\mathrm{rel}};L,M)
=
\frac{
M^{2}\bigl[(L+1)^{2} + (L\cos\phi_{\mathrm{rel}} + 1)^{2}\bigr]^{2}
}{
64\,L^{2}\, |f(0,\phi_{\mathrm{rel}})|^{2}
}
\label{eq:Psi3_def}
\end{equation}
characterizes the multiplicative degradation caused by AP-side misalignment. The composite term inside the brackets reflects the distortion of the near-field boundary due to the effective AP rotation, whereas the factor $|f(0,\phi_{\mathrm{rel}})|^{2}$ accounts for the element-pattern loss induced by the rotated boresight.

As expected, when $\phi_{\mathrm{rel}}=0$, we have $|f(0,0)|^{2}=1$ and $\Psi_{3}(0;L,M)={M^{2}(L+1)^{4}}/({16L^{2}}),$ which exactly recovers the aligned mobile limit in~\eqref{eq:bw_mobile_ratio} on the top of the next page. 

By inverting~\eqref{eq:bw_AP_final} and substituting~\eqref{eq:Psi3_def}, the minimum transmit power required to support a target bandwidth $B$ while operating \emph{exclusively} in the far field is in~\eqref{equ:Pt3}. 
\begin{figure*}
    \begin{equation}
P_{\mathrm{t},\mathrm{dBm}}^{\mathrm{AP}1}\left( \phi _{\mathrm{rel}} \right) =64+\mathsf{SNR}_{\mathrm{th},\mathrm{dB}}+N_{\mathrm{F},\mathrm{dB}}+10\log _{10}\left( kT \right) +10\log _{10}\left( B_{\mathrm{Mobile},\mathrm{AP}}^{(\max\mathrm{)}}(\phi _{\mathrm{rel}}) \right) +10\log _{10}\left( \Psi _3\left( \phi _{\mathrm{rel}};L,M \right) \right) .
\label{equ:Pt3}
\end{equation}
\hrulefill
\end{figure*}

\subsection{Scenario 4: Relative Angular Variation With Two Angles}
\label{sec:scenario4}

We now extend Scenario~3 to a two-dimensional motion-induced \emph{relative} angular variation.
In \textit{Scenario~4}, shown in Fig.~\ref{fig:scenario4}, the AP is fixed and the UE does not self-rotate; instead, UE motion induces relative angular offsets in both the vertical and horizontal planes, denoted by $(\theta_{\mathrm{rel}},\phi_{\mathrm{rel}})$.
While this scenario may be less representative of typical handheld UEs, it serves as a useful limiting case and can also model stabilized platforms (e.g., airborne nodes) that approximately maintain the array orthogonal to the LoS.

For the UPA-UPA case with two relative angles $(\theta_{\mathrm{rel}},\phi_{\mathrm{rel}})$, the corresponding near-field distance is given in~\cite{zhang2025misalignedTWC} as
\begin{equation}
\begin{split}
&d_{\mathrm{F}}^{(\mathrm{AP)}}(\theta_{\mathrm{rel}},\phi _{\mathrm{rel}})=\frac{2\left( D_1\cos \theta_{\mathrm{rel}}+D_2 \right) ^2}{\lambda}
\\
&+\frac{2\left( D_1(\cos \phi_{\mathrm{rel}}+|\sin \theta _{\mathrm{rel}}\sin \phi_{\mathrm{rel}}|)+D_2 \right) ^2}{\lambda}.
\end{split}
\label{eq:dF_AP_two_angles}
\end{equation}

Enforcing \textit{Condition~1}, $d_{\min} \ge d_{\mathrm{F}}^{(\mathrm{AP})}$, and substituting $D_{1}=L D_{2}$ with $L = D_{1}/D_{2}$ yield
\begin{equation}
F_{\mathrm{AP}}(L,\theta_{\mathrm{rel}},\phi_{\mathrm{rel}})
\, D_{2}^{2}
\le
\frac{ \lambda d_{\min}}{2},
\label{eq:cond1_AP_two_LM}
\end{equation}
where
\begin{equation}
\begin{split}
F_{\mathrm{AP}}(L,\theta _{\mathrm{rel}},\phi _{\mathrm{rel}})=\bigl( L\cos \theta _{\mathrm{rel}}+1 \bigr) ^2+
\\
\bigl( L(\cos \phi _{\mathrm{rel}}+|\sin \theta _{\mathrm{rel}}\sin \phi _{\mathrm{rel}}|)+1 \bigr) ^2.
\end{split}
\label{eq:F_AP_two_def}
\end{equation}

The UE remains aligned, so $G_{\mathrm{UE}}(0,0)=N_{2}^{2} G_{0}.$ The AP gain now depends on both rotation angles
\begin{equation}
G_{\mathrm{AP}}(\theta_{\mathrm{rel}},\phi_{\mathrm{rel}})
=
N_{1}^{2} G_{0}
\, |f(\theta_{\mathrm{rel}},\phi_{\mathrm{rel}})|^{2}.
\label{eq:G_AP_two_angles}
\end{equation}

Substituting \eqref{eq:G_AP_two_angles} into the SNR expression and imposing \textit{Condition~2}, $\mathsf{SNR}(d_{\max}) \ge \mathsf{SNR}_{\mathrm{th}}$, we obtain

\begin{equation}
D_{1} D_{2}
\ge
\frac{\lambda \pi d_{\max}}{G_{0}|f(\theta_{\mathrm{rel}},\phi_{\mathrm{rel}})|}
\sqrt{\frac{N_{\mathrm{F}}kTB}{P_{\mathrm{t}}}}
\,10^{\frac{\mathsf{SNR}_{\mathrm{th,dB}}}{20}}.
\label{eq:cond2_AP_two}
\end{equation}

Using $D_{1}=L D_{2}$ and $d_{\max}= M d_{\min}$ gives
\begin{equation}
L D_{2}^{2}
\ge
\frac{\lambda \pi M d_{\min}}{G_{0}|f(\theta_{\mathrm{rel}},\phi_{\mathrm{rel}})|}
\sqrt{\frac{N_{\mathrm{F}}kTB}{P_{\mathrm{t}}}}
\,10^{\frac{\mathsf{SNR}_{\mathrm{th,dB}}}{20}}.
\label{eq:cond2_AP_two_LM}
\end{equation}

Combining \eqref{eq:cond1_AP_two_LM} and \eqref{eq:cond2_AP_two_LM} and eliminating $D_{2}^{2}$ in the same manner as Scenarios~0--3, we obtain the maximum achievable bandwidth under two AP rotation angles
\begin{equation}
\begin{split}
B_{\mathrm{Mobile},\mathrm{AP}2}^{(\max\mathrm{)}}(\theta _{\mathrm{rel}},\phi _{\mathrm{rel}})&\le \frac{L^2G_{0}^{2}|f(\theta _{\mathrm{rel}},\phi _{\mathrm{rel}})|^2}{4\pi ^2kTM^2F_{\mathrm{AP}}^{2}(L,\theta _{\mathrm{rel}},\phi _{\mathrm{rel}})}
\\
&\times \,10^{\frac{P_{\mathrm{t},\mathrm{dBm}}-\mathsf{SNR}_{\mathrm{th},\mathrm{dB}}-N_{\mathrm{F},\mathrm{dB}}-30}{10}}.
\end{split}
\label{eq:bw_AP_two_angles}
\end{equation}

By expressing the maximum bandwidth \eqref{eq:bw_AP_two_angles} in Scenario~4 relative to the stationary limit~\eqref{eq:bw_final_stationary_new}, we obtain
\begin{equation}
B_{\mathrm{Mobile,AP2}}^{(\max)}(\theta_{\mathrm{rel}},\phi_{\mathrm{rel}})
=
\frac{
B_{\mathrm{Stationary}}^{(\max)}
}{
\Psi_{4}(\theta_{\mathrm{rel}},\phi_{\mathrm{rel}};L,M)
},
\label{eq:bw_AP2_ratio}
\end{equation}
where
\begin{equation}
\Psi_{4}(\theta_{\mathrm{rel}},\phi_{\mathrm{rel}};L,M)
=
\frac{
M^{2}\, F_{\mathrm{AP}}^{2}(L,\theta_{\mathrm{rel}},\phi_{\mathrm{rel}})
}{
64\,L^{2}\, |f(\theta_{\mathrm{rel}},\phi_{\mathrm{rel}})|^{2}
},
\label{eq:Psi4_def}
\end{equation}
quantifies the multiplicative degradation caused by two-dimensional AP-side misalignment. 

In the special case $(\theta_{\mathrm{rel}},\phi_{\mathrm{rel}})=(0,0)$, we have 
$F_{\mathrm{AP}}(L,0,0)=2{(L+1)^{2}}$ and $|f(0,0)|^{2}=1$, which yields $\Psi_{4}(0,0;L,M)={M^{2}(L+1)^{4}}/({16L^{2}}),$ thus recovering both the aligned AP case in Scenario~3 and the baseline mobile limit in~\eqref{eq:bw_mobile_ratio}.

By inverting~\eqref{eq:bw_AP_two_angles} and substituting~\eqref{eq:Psi4_def}, the minimum transmit power required to support a target bandwidth $B$ while operating \emph{exclusively} in the far field is 
\begin{equation}
    \begin{split}
        &P_{\mathrm{t},\mathrm{dBm}}^{\mathrm{AP}2}\left( \theta _{\mathrm{rel}},\phi _{\mathrm{rel}} \right) =64+\mathsf{SNR}_{\mathrm{th},\mathrm{dB}}+N_{\mathrm{F},\mathrm{dB}}\\&+10\log _{10}\left( kT \right) +10\log _{10}\left( B_{\mathrm{Mobile},\mathrm{AP}}^{(\max)}(\theta _{\mathrm{rel}},\phi _{\mathrm{rel}}) \right) \\&+10\log _{10}\left( \Psi _4\left( \theta _{\mathrm{rel}},\phi _{\mathrm{rel}};L,M \right) \right).
    \end{split}
\end{equation}

\section{Numerical Results}
\label{sec:numerical}
In this section, we numerically evaluate the far-field feasibility limits for stationary and mobile THz systems established through the proof-by-contradiction analysis in Section~\ref{sec:analysis}. 

To model element directivity in a realistic yet tractable manner, we adopt a symmetric cosine-pattern element with field response $f(\theta,\phi)=\cos^{q}\theta \cos^{q}\phi$, where the non-negative parameter $q$ controls the directivity in both the vertical and horizontal planes~\cite{brar2022dual}. Unless otherwise stated, we use $q=1$. This corresponds to a moderately directive element with mild angular selectivity. In this case, the normalization constant is $G_0=6$, while the general expression and its derivation are given in Appendix~\ref{app:cosine_element}. The system temperature is set to $T=290~\mathrm{K}$, the Boltzmann constant is $k_{\mathrm{B}}=1.38\times10^{-23}~\mathrm{J/K}$, and the target SNR threshold is fixed to $30~\mathrm{dB}$\footnote{This value should not be interpreted as a typical cell-edge SNR, but rather as a \emph{design-margin} requirement for a LoS-dominated THz link intended to support very high spectral efficiency and to leave headroom for practical impairments (e.g., tracking errors, hardware nonidealities, weather-related attenuation, and occasional NLoS/blockage losses). In conventional cellular systems, the cell-edge SNR is often only a few dB due to fading, NLoS conditions, and antenna misalignment. Hence, the corresponding \emph{ideal LoS} SNR budget needs to be substantially higher to maintain reliable operation under adverse propagation~effects.}, which is sufficient for reliable high-rate THz operation~\cite{sen2022terahertz}. The uplink transmit power of the mobile node is set to $P_{\mathrm{t,dBm}}=23$~dBm, and the receiver noise figure is $N_\mathrm{F}=10$~dB.

\subsection{Stationary THz Link}
\label{sec:stationary_results}

\begin{figure*}[!t]
    \centering
    \subfloat[Effect of the transmit power]{
        \includegraphics[width=0.32\textwidth]{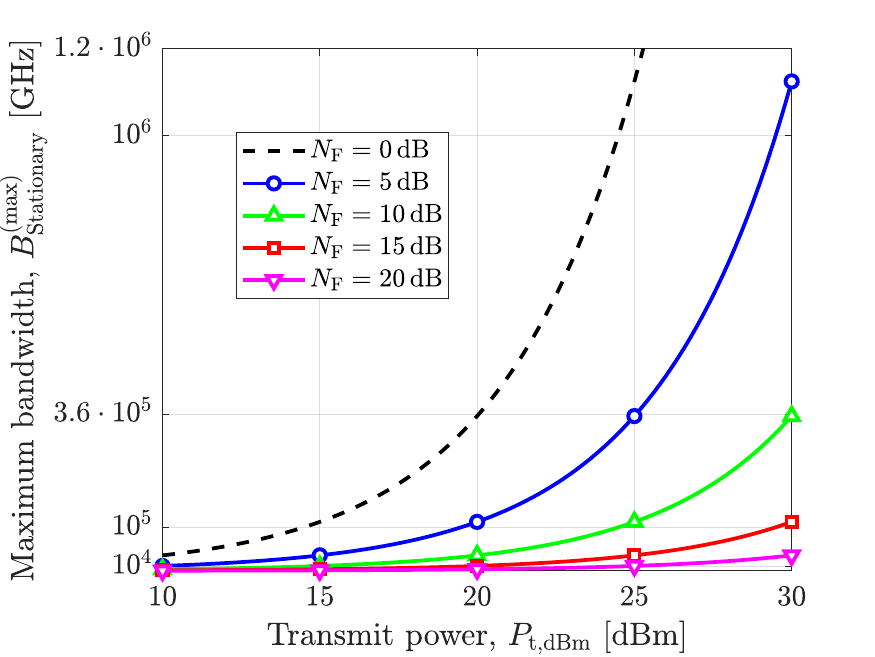}%
        \label{fig:stationary_Pt}
    }
    \hspace{-3mm}
    \subfloat[Effect of the SNR threshold]{
        \includegraphics[width=0.32\textwidth]{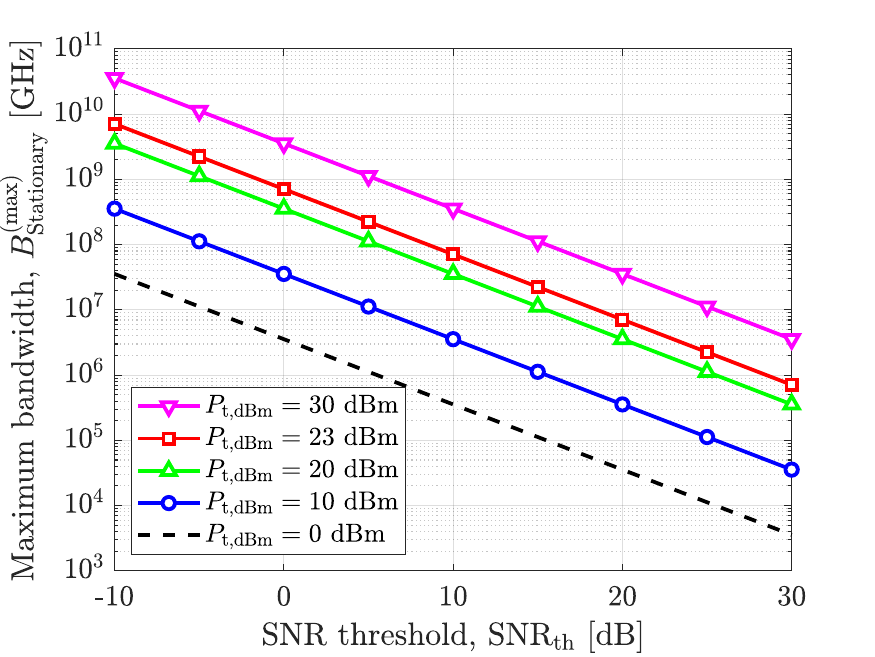}%
        \label{fig:stationary_SNR}
    }
    \hspace{-3mm}
    \subfloat[Effect of the central frequency]{
        \includegraphics[width=0.32\textwidth]{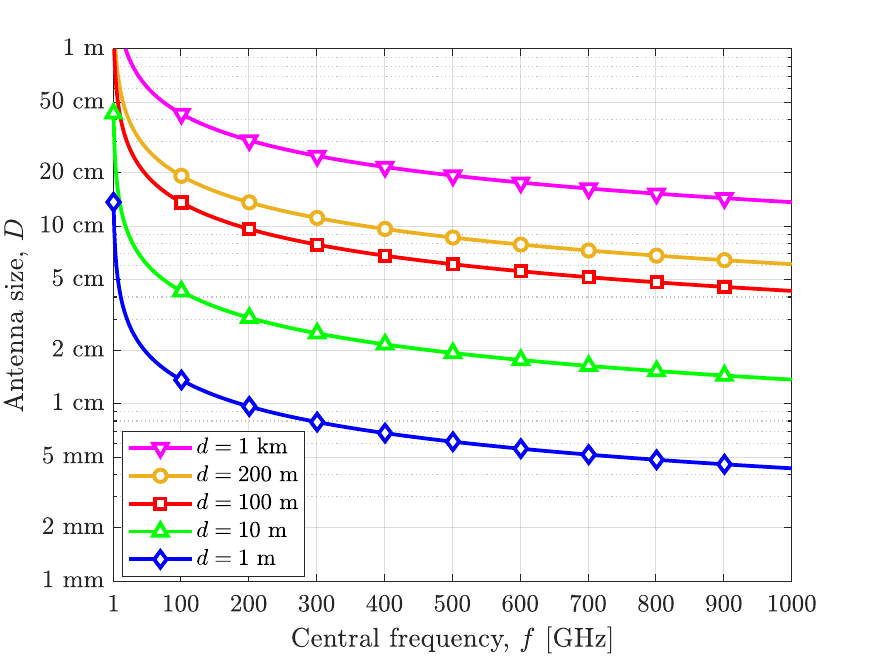}%
        \label{fig:stationary_freq}
    }
    \caption{The maximum achievable bandwidth for the far-field stationary THz link and the corresponding antenna sizes.}
    \label{fig:stationary_all}
    \vspace{-0.4cm}
\end{figure*}

We first consider a stationary THz link designed to operate exclusively in the far field. Fig.~\ref{fig:stationary_all} shows the maximum achievable bandwidth $B^{(\max)}_{\mathrm{Stationary}}$ and the corresponding antenna apertures under different system parameters. Specifically, Fig.~\ref{fig:stationary_Pt} plots $B^{(\max)}_{\mathrm{Stationary}}$ versus the transmit power $P_{\mathrm{t,dBm}}$ for different receiver noise figures. The dashed curve with $\mathrm{NF}=0$~dB serves as an idealized reference corresponding to a noise-free receiver. As expected, increasing the transmit power or reducing the noise figure improves the achievable bandwidth. Notably, even for practical transmit powers on the order of tens of milliwatts (i.e., $P_{\mathrm{t,dBm}}\in[10,30]$~dBm), the achievable bandwidth readily reaches the terahertz range, consistent with experimentally demonstrated THz transmit powers~\cite{sen2020teranova}. Fig.~\ref{fig:stationary_SNR} further shows $B^{(\max)}_{\mathrm{Stationary}}$ as a function of the SNR threshold $\mathsf{SNR}_{\mathrm{th}}$ for different transmit power levels. Relaxing the SNR requirement substantially increases the achievable bandwidth, further confirming that, in stationary scenarios, neither practical transmit-power limitations nor stringent SNR targets fundamentally force the system into near-field operation. In fact, over a broad range of practical parameters (e.g., $\mathsf{SNR}_{\mathrm{th}}\leq 25$~dB), the achievable bandwidth can even exceed the carrier frequency itself, which indicates that the far-field constraint does not create an intrinsic bandwidth bottleneck in stationary scenarios.

While the achievable bandwidth in Fig.~\ref{fig:stationary_Pt} and Fig.~\ref{fig:stationary_SNR} is independent of the communication distance and wavelength, these parameters directly determine the required antenna apertures. This is illustrated in Fig.~\ref{fig:stationary_freq}, which shows the antenna size $D$ required at both the transmitter and receiver to maintain far-field operation while achieving $B^{(\max)}_{\mathrm{Stationary}}$. As the carrier frequency increases, the required aperture decreases due to the combined effects of the array gain scaling with $1/\lambda^2$ and the near-field boundary scaling with $1/\lambda$. Conversely, larger communication distances require larger apertures to compensate for the increased free-space path loss.

Despite these trends, the required antenna sizes remain practically feasible in representative THz settings. For example, at $300$~GHz and a link distance of $200$~m, an aperture of about $10$~cm $\times$ $10$~cm is sufficient to sustain the maximum achievable bandwidth while remaining in the far field. At higher carrier frequencies, the required aperture becomes even smaller. In extreme low-frequency regimes, such as $f=1$~GHz and $d=100$~m, the required aperture may exceed $1$~m $\times$ $1$~m. This does not indicate that far-field operation is infeasible; rather, it reflects that we intentionally maximize the bandwidth, which can become unrealistically large in stationary scenarios\footnote{For a given aperture, the attainable bandwidth can be directly computed from the far-field feasibility condition in~\eqref{eq:cond2_N}. In the same low-frequency example, even a much smaller aperture, e.g., $20$~cm $\times$ $20$~cm, already supports a maximum bandwidth on the order of the carrier frequency (about $1$~GHz), which is sufficient for practical broadband operation. Therefore, the large apertures in Fig.~\ref{fig:stationary_freq} should be interpreted as those required to attain the absolute theoretical maximum, whereas practically relevant bandwidths can be achieved with substantially smaller and physically realizable arrays.}.

Overall, Fig.~\ref{fig:stationary_all} shows that a stationary THz link can be designed to operate exclusively in the far field without sacrificing bandwidth, using realistic transmit powers, receiver noise figures, and physically realizable antenna sizes.

\begin{figure*}[!t]
    \centering
    \subfloat[Effect of mobility coefficient]{
        \includegraphics[width=0.32\textwidth]{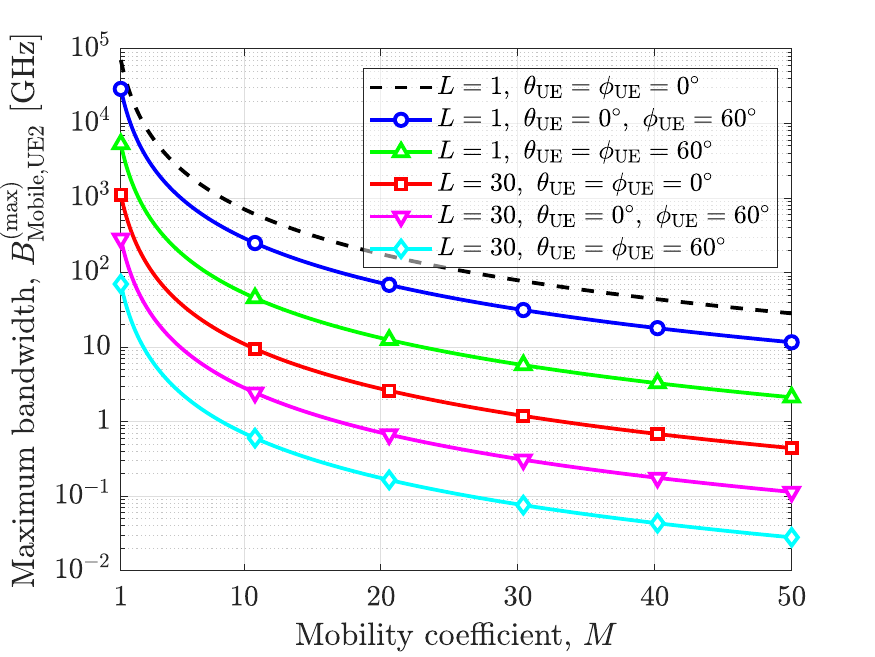}%
        \label{fig:Fig2a_UErot_M_Bmax}
    }\hspace{-3mm}
    \subfloat[Effect of antenna inequality coefficient]{
        \includegraphics[width=0.32\textwidth]{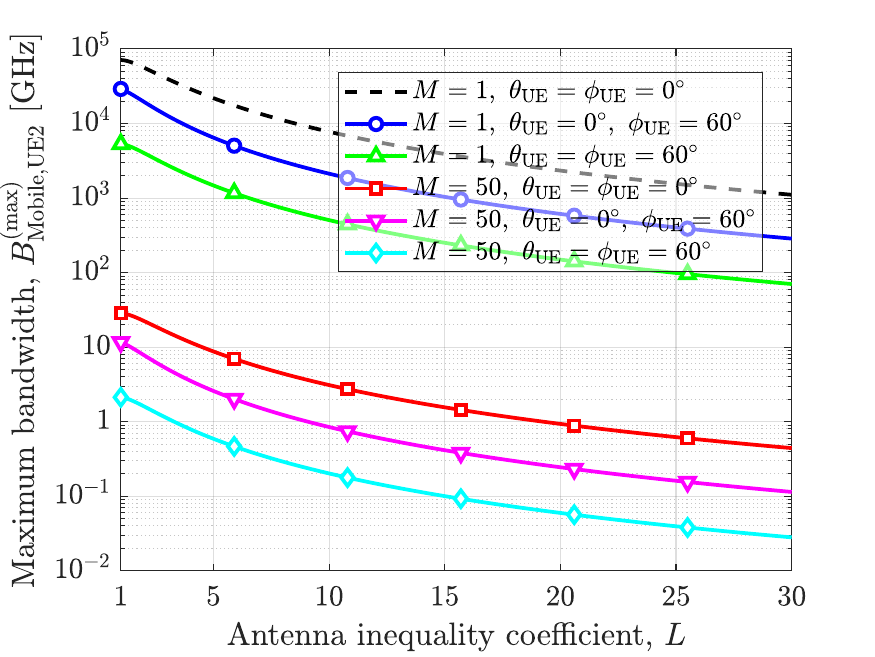}%
        \label{fig:Fig3a_UErot_L_Bmax}
    }\hspace{-3mm}
    \subfloat[Sanity check of the aperture sizes]{
        \includegraphics[width=0.32\textwidth]{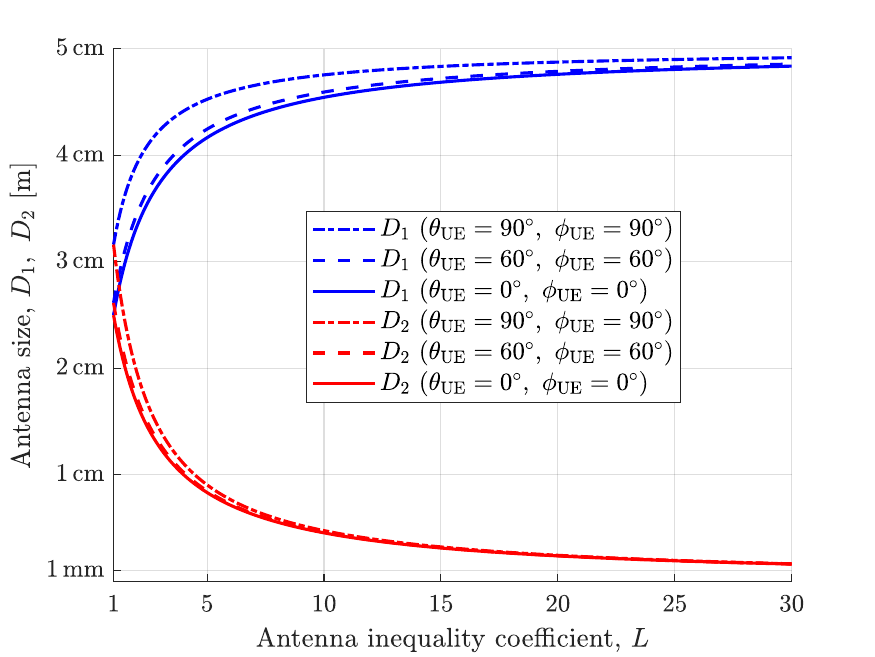}%
        \label{fig:Fig6a_UE_sanity_check_THz}
    }

    \caption{Maximum achievable bandwidth under UE rotation and mobility.}
    \label{fig:Fig2_all}
    \vspace{-0.3cm}
\end{figure*}

\subsection{Mobile THz Link Under UE Rotation}
\label{sec:mobile_results}
\vspace{-1mm}

We next switch from stationary to mobile THz communications in Fig.~\ref{fig:Fig2_all}, where the far-field feasibility must be maintained over a \emph{range} of link distances rather than a single value. This introduces three practical constraints: (i) the SNR requirement must be satisfied at the largest distance $d_{\max}$, whereas the array sizes are upper-bounded by the far-field constraint at the smallest distance $d_{\min}$, which we capture via the mobility coefficient $M \triangleq d_{\max}/d_{\min}$ (the worst-case link budget); (ii) the antenna apertures at the AP and UE are generally asymmetric, which we model via the antenna inequality coefficient $L \triangleq D_1/D_2$; and (iii) the uplink transmit power of a battery-powered UE is typically more limited than that of a stationary backhaul transmitter. In addition to these mobility-induced constraints, we explicitly incorporate UE rotation, characterized by the rotation angles $(\theta_{\mathrm{UE}},\phi_{\mathrm{UE}})$.

Fig.~\ref{fig:Fig2a_UErot_M_Bmax} shows the maximum achievable bandwidth as a function of the mobility coefficient $M$ under different antenna inequality levels $L$ and UE rotation angles. The overall trend follows intuition: $B^{(\max)}_{\mathrm{Mobile,UE2}}$ decreases monotonically with $M$ since a larger mobility range tightens the simultaneous satisfaction of the SNR constraint at $d_{\max}$ and the far-field constraint at $d_{\min}$. For the aligned case $(\theta_{\mathrm{UE}},\phi_{\mathrm{UE}})=(0^\circ,0^\circ)$, the bandwidth remains on the order of several terahertz at small $M$ (e.g., $M\leq 10$) and drops to the gigahertz range at realistic mobility levels (e.g., $M\approx 40$--$50$) when symmetric apertures are used (i.e., $L=1$). 
Once UE rotation is introduced, the bandwidth is reduced by \emph{orders of magnitude} compared to the aligned baseline, and this degradation becomes more pronounced as $M$ increases, since mobility and rotation jointly shrink the available link margin.

Fig.~\ref{fig:Fig3a_UErot_L_Bmax} depicts $B^{(\max)}_{\mathrm{Mobile,UE2}}$ versus the antenna inequality coefficient $L$ for representative mobility levels. As expected, increasing $L$ reduces the achievable bandwidth due to the smaller UE aperture $D_2$ (and hence reduced UE array gain). When there is no mobility ($M=1$), the system can still support extremely large bandwidths even for strong inequality (e.g., $L=30$), consistent with the stationary-like behavior. In contrast, under realistic mobility (e.g., $M=50$), the bandwidth becomes severely constrained for practical values of $L$ (e.g., $L\approx 20$--$30$), and UE rotation further tightens these feasibility limits, which leads to an additional reduction of several orders of magnitude relative to the aligned case.

Finally, Fig.~\ref{fig:Fig6a_UE_sanity_check_THz} provides a sanity check for the resulting antenna apertures $(D_1,D_2)$ implied by the far-field feasibility conditions. The obtained sizes remain physically realizable across the considered parameters. Importantly, UE rotation does not introduce any unrealistic aperture requirements, while it primarily manifests as a gain penalty that directly reduces $B^{(\max)}_{\mathrm{Mobile,UE2}}$ in Fig.~\ref{fig:Fig2a_UErot_M_Bmax} and Fig.~\ref{fig:Fig3a_UErot_L_Bmax}. 

Overall, Fig.~\ref{fig:Fig2_all} demonstrates that \textbf{while mobility and antenna inequality already impose stringent far-field feasibility limits, incorporating UE rotation makes the bandwidth constraints substantially more severe in practically relevant regimes.}

\begin{figure*}[!t]
    \centering


    \subfloat[Effect of mobility coefficient]{
    \includegraphics[width=0.32\textwidth]{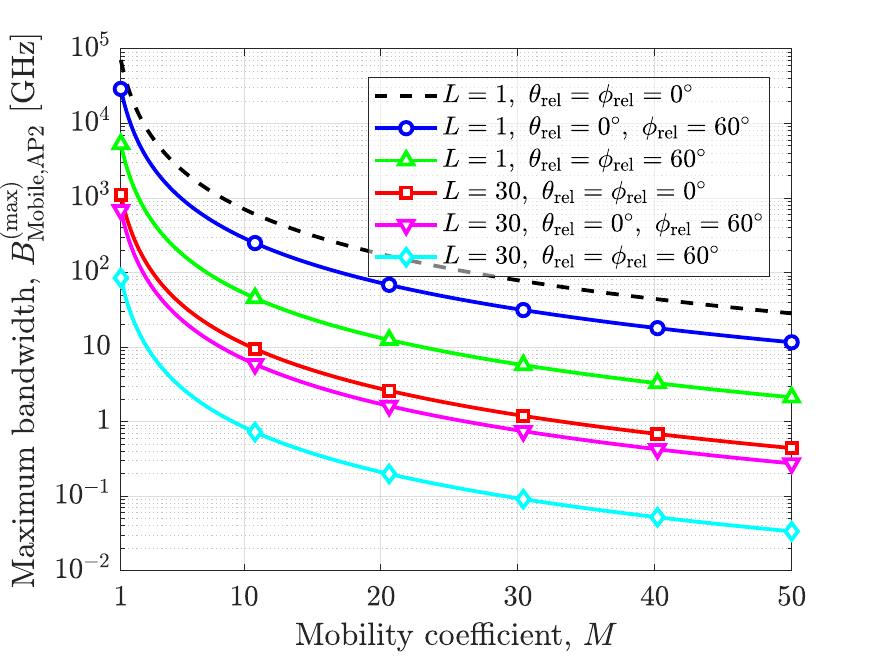}%
    \label{fig:Fig2b_AProt_M_Bmax}
    }\hspace{-3mm}
    \subfloat[Effect of antenna inequality coefficient]{
        \includegraphics[width=0.32\textwidth]{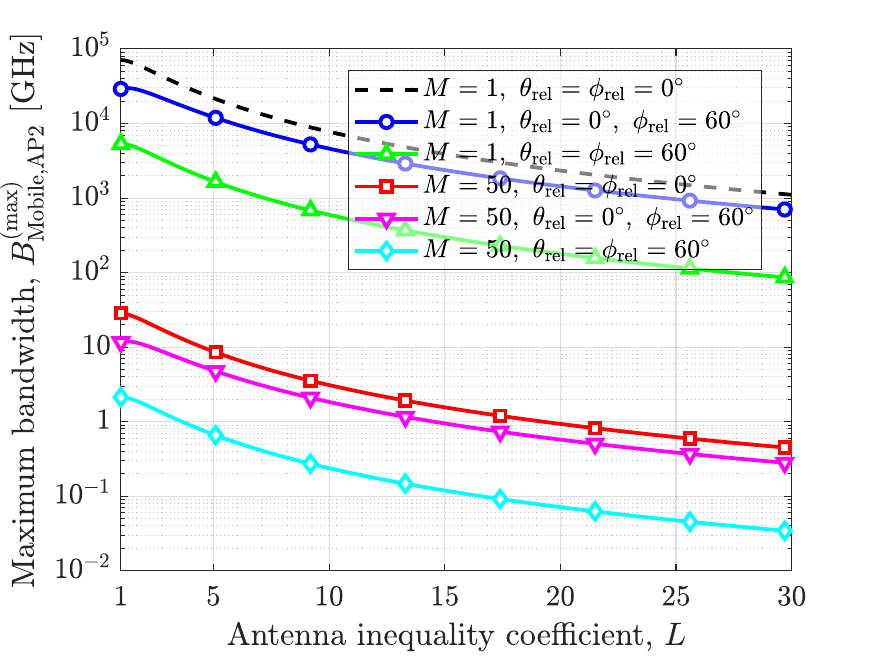}%
        \label{fig:Fig3b_AProt_L_Bmax}
    }\hspace{-3mm}
    \subfloat[Sanity check of the aperture sizes]{
    \includegraphics[width=0.32\textwidth]{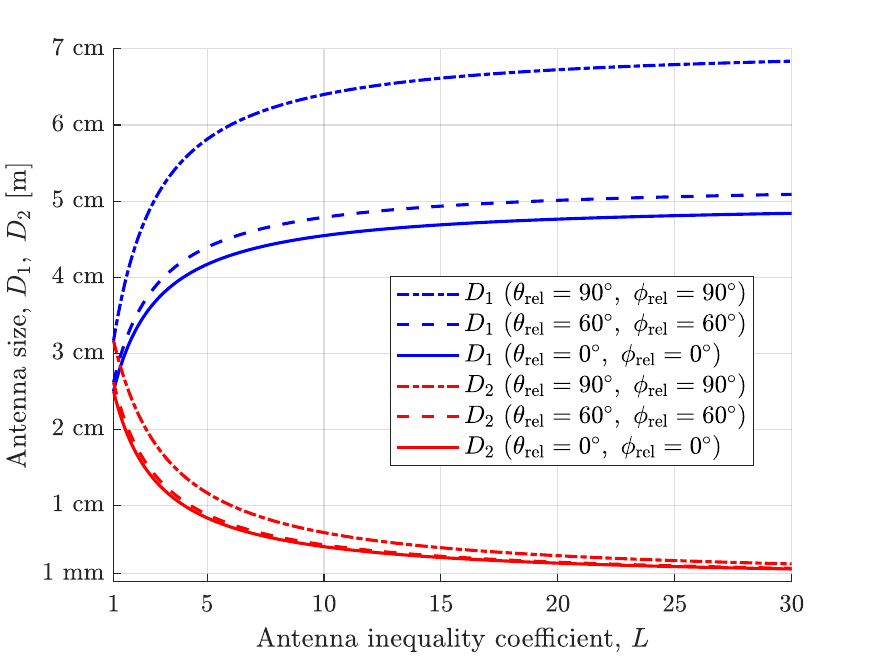}%
    \label{fig:Fig7a_AP_sanity_check_THz}
    }

    \caption{Maximum achievable bandwidth under direction variation and mobility.}
    \label{fig:Fig3_all}
    \vspace{-0.5cm}
\end{figure*}

\subsection{Mobile THz Link Under Direction Variation}
\label{sec:relative_rotation_results}

We next model the orientation mismatch through \emph{relative} rotation between the AP and UE, characterized by the relative angles $(\theta_{\mathrm{rel}},\phi_{\mathrm{rel}})$. While the mobility- and asymmetry-induced constraints captured by $M$ and $L$ remain unchanged, \textit{direction variation} affects the feasibility limits through the same element-pattern gain penalty. As evidenced by Figs.~\ref{fig:Fig2_all} and~\ref{fig:Fig3_all}, UE rotation and direction variation lead to bandwidth limitations of comparable order. Hence, the dominant factor is the magnitude of the resulting misalignment rather than whether it comes from UE rotation or direction variation.

Fig.~\ref{fig:Fig2b_AProt_M_Bmax} shows the maximum achievable bandwidth $B^{(\max)}_{\mathrm{Mobile,AP2}}$ as a function of the mobility coefficient $M$ for different antenna inequality levels $L$ and direction variation angles. Similar to Fig.~\ref{fig:Fig2a_UErot_M_Bmax}, $B^{(\max)}_{\mathrm{Mobile,AP2}}$ decreases monotonically with $M$. Moreover, non-zero direction variation causes an orders-of-magnitude degradation compared to the aligned case $(\theta_{\mathrm{rel}},\phi_{\mathrm{rel}})=(0^\circ,0^\circ)$, and this degradation becomes increasingly pronounced for realistic mobility levels (e.g., $M\approx 40$--$50$), where the feasibility margin is already limited.

Fig.~\ref{fig:Fig3b_AProt_L_Bmax} depicts $B^{(\max)}_{\mathrm{Mobile,AP2}}$ versus the antenna inequality coefficient $L$ for representative mobility levels. Increasing $L$ reduces the available bandwidth due to the smaller UE aperture $D_2$ and the associated loss in array gain. When there is no mobility ($M=1$), the system can still sustain large bandwidths even for substantial inequality (e.g., $L=30$). In contrast, under realistic mobility (e.g., $M=50$), the achievable bandwidth becomes severely constrained for practical values of $L$, and direction variation further tightens these~limits.

Finally, Fig.~\ref{fig:Fig7a_AP_sanity_check_THz} provides a sanity check for the implied aperture sizes $(D_1,D_2)$ under representative direction variations. The resulting sizes remain physically realizable and preserve the expected scaling with $L$, which demonstrates that the bandwidth degradation in Fig.~\ref{fig:Fig2b_AProt_M_Bmax} and Fig.~\ref{fig:Fig3b_AProt_L_Bmax} is primarily driven by the misalignment-induced gain penalty rather than by unrealistic aperture requirements. 

Overall, Fig.~\ref{fig:Fig3_all} demonstrates that \textbf{under mobility and antenna inequality, orientation mismatch---here modeled by direction variation---can substantially tighten the far-field achievable bandwidth.}

\begin{figure*}[!t]
    \centering

    \subfloat[sub-6\,GHz bandwidth]{
        \includegraphics[width=0.34\textwidth]{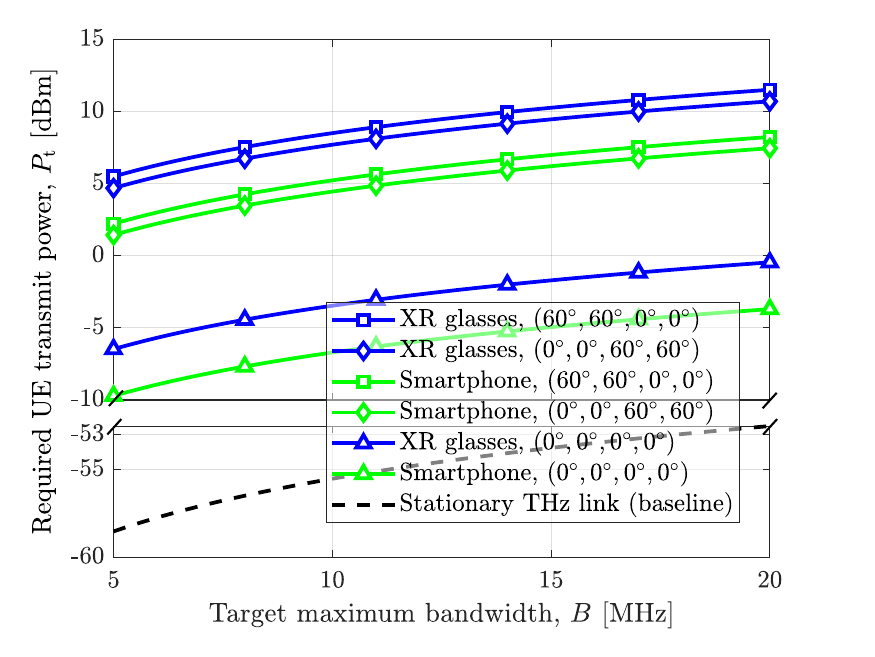}%
        \label{fig:Fig4a_UE_required_power_sub6}
    }\hspace{-7mm}
    \subfloat[mmWave bandwidth]{
        \includegraphics[width=0.34\textwidth]{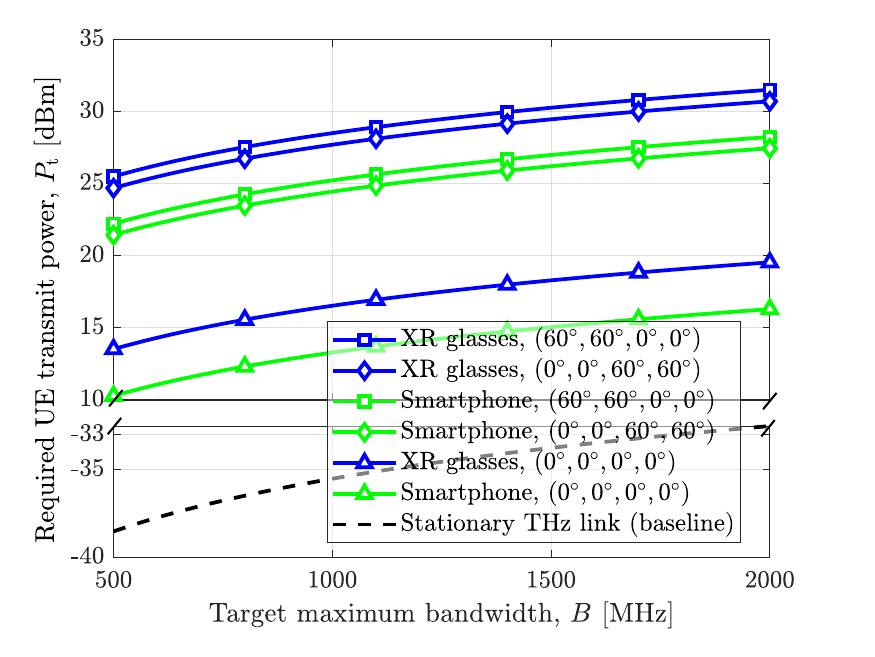}%
        \label{fig:Fig4b_UE_required_power_mmWave}  
    }\hspace{-7mm}
    \subfloat[THz bandwidth]{
        \includegraphics[width=0.34\textwidth]{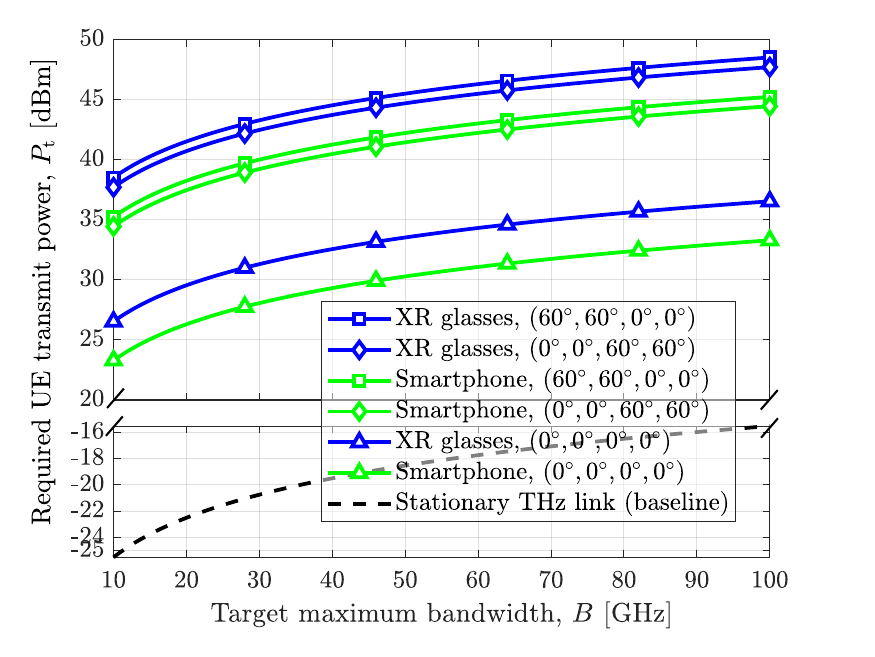}%
        \label{fig:Fig4c_UE_required_power_THz}
    }

    \caption{Required UE transmit power $P_{\mathrm{t}}$ as a function of the target maximum bandwidth $B$ for (a) sub-6\,GHz, (b) mmWave, and (c) THz bands under different UE rotation and direction variation conditions. The four-tuple $(\theta_{\mathrm{UE}},\phi_{\mathrm{UE}},\theta_{\mathrm{rel}},\phi_{\mathrm{rel}})$ specifies the UE rotation and direction variation angles.}
    \label{fig:Fig4_all}
\end{figure*}

\subsection{Do Mobile THz Links Have to Operate in the Near Field?}
\label{sec:nearfield_inevitability}

We finally quantify whether mobile links can remain exclusively in the far field while meeting a target bandwidth requirement. Fig.~\ref{fig:Fig4_all} shows the \emph{required} UE transmit power $P_{\mathrm{t,dBm}}$ as a function of the target maximum bandwidth $B$ across three carrier-frequency regimes: sub-6~GHz, mmWave, and THz. We set the target SNR threshold to $\mathsf{SNR}_{\mathrm{th}}=20$~dB to retain a reasonable link margin under practical propagation impairments. We focus on a typical indoor-access scenario with $M=50$ (from $0.5$~m to $25$~m~\cite{petrov2023near}) and consider two prospective UE form factors, namely, a smartphone ($L\approx 20$) and XR glasses ($L\approx 30$). The dashed curve in each subplot corresponds to the stationary-link baseline. The four-tuple $(\theta_{\mathrm{UE}},\phi_{\mathrm{UE}},\theta_{\mathrm{rel}},\phi_{\mathrm{rel}})$ specifies the UE rotation and direction variation angles used in the evaluation.

From Fig.~\ref{fig:Fig4a_UE_required_power_sub6}, in the sub-6~GHz regime the required UE transmit power remains modest over the considered bandwidth range. For instance, at $B=20$~MHz, the required $P_{\mathrm{t,dBm}}$ is approximately $0$~dBm for XR glasses in the aligned case and about $-3$~dBm for the smartphone, whereas incorporating rotation increases the required power to roughly $10$--$12$~dBm (XR glasses) and $7$--$8$~dBm (smartphone). \textit{Hence, even with rotation-induced misalignment, far-field operation does not demand excessive UE power in this regime.}

The mmWave case in Fig.~\ref{fig:Fig4b_UE_required_power_mmWave} exhibits the same monotonic trend that larger target bandwidth requires higher transmit power, but the absolute power levels increase noticeably. At the lower end ($B=400$~MHz, as the maximum bandwidth in 5G-grade FR2 systems~\cite{3gpp38104}), required $P_{\mathrm{t,dBm}}$ is around $12$~dBm for the aligned smartphone case and about $15$~dBm for the aligned XR-glasses case, while rotation raises the requirement to approximately $23$~dBm (smartphone) and $26$--$27$~dBm (XR glasses). At the upper end of the shown range ($B=2$~GHz, as the maximum bandwidth for 5G-Advanced FR2--2 systems~\cite{3gpp38104}), the required $P_{\mathrm{t,dBm}}$ is around $16$--$17$~dBm for the aligned smartphone case and about $19$--$20$~dBm for the aligned XR-glasses case, while rotation raises the requirement to approximately $27$--$28$~dBm (smartphone) and $30$--$32$~dBm (XR glasses). \textit{This indicates that mmWave links can remain far-field-feasible at \emph{moderate} bandwidths within typical UE power budgets, whereas pushing towards multi-GHz bandwidths may already approach or exceed practical limits, particularly for smaller wearable form factors.}

The most critical regime is THz in Fig.~\ref{fig:Fig4c_UE_required_power_THz}. Here, the required UE transmit power grows rapidly with the target bandwidth and becomes prohibitively large under mobility and orientation mismatch. For example, already at $B=10$~GHz, the required power for every considered case exceeds, often by far, $23$~dBm (200~mW). \emph{Hence, even for moderate (for THz frequencies) bandwidth, the required UE transmit power in a mobile far-field only regime becomes greater than most of the existing mobile THz hardware systems can deliver~\cite{jornetProcIEEE}.} The minimal level of $P_{\mathrm{t,dBm}}$ also grows rapidly with bandwidth. At $B=100$~GHz, the aligned cases already require about $36$~dBm (XR glasses) and $33$~dBm (smartphone), whereas incorporating either UE rotation or direction variation increases the requirement to approximately $48$~dBm (XR glasses) and $45$~dBm (smartphone), i.e., an additional penalty on the order of $10$~dB. \textit{These power levels are far above what is typically available for battery-powered mobile devices~\cite{john2020broadband}.}

Overall, Fig.~\ref{fig:Fig4_all} reveals a clear contrast across frequency regimes: while sub-6~GHz links \emph{can} remain in the far field without demanding excessive UE transmit power (even under rotation), and mmWave links \emph{may still be feasible} for moderate target bandwidths, mobile THz links targeting tens of gigahertz of bandwidth would \emph{require unrealistically high UE transmit power} if constrained to operate exclusively in the far field. \textbf{Therefore, for practically relevant mobile THz access scenarios, operating exclusively in the far field is essentially infeasible, and near-field operation becomes unavoidable.}

\section{Conclusion}

This paper investigated whether a mobile THz link can operate \emph{exclusively} in the radiative far field while remaining broadband. To this end, we developed a proof-by-contradiction feasibility framework that simultaneously enforces the Fraunhofer-based near-field condition and a worst-case SNR requirement over a mobility-induced distance range. By adopting a generalized planar array-gain model, we derived closed-form expressions for the maximum far-field-feasible bandwidth in both stationary and mobile settings. We finally quantified how practical deployment configurations and antenna array misalignment can affect the feasibility limits.

\emph{The analysis and numerical results lead to two main insights.} First, \emph{stationary} THz links \emph{can} be designed to remain far-field-only without sacrificing bandwidth: the far-field constraint does not impose an intrinsic bandwidth bottleneck as long as physically realizable apertures and reasonable link budgets are used. Second, mobility fundamentally changes the feasibility landscape. As a result, for practically relevant \emph{mobile} THz access scenarios, \emph{operating exclusively in the far field becomes essentially infeasible} when targeting the wide bandwidths that motivate THz communications in the first place. A cross-band evaluation further showed that far-field-only operation remains substantially more attainable at sub-6~GHz and, to a large extent, at mmWave for moderate bandwidths, whereas the THz regime is markedly more sensitive to mobility, form-factor constraints, and orientation mismatch. This suggests that, in contrast to lower frequencies, \emph{practical mobile THz systems must be always designed as near-field-aware}, rather than relying on far-field plane-wave assumptions.

\appendices
\renewcommand{\theequation}{\thesection-\arabic{equation}}
\setcounter{equation}{0}

\section{Derivation of the Cosine-Pattern Element}
\label{app:cosine_element}

In this appendix, we derive the closed-form expression of the element-pattern normalization constant $G_0$ for the symmetric cosine-pattern element used in Section~\ref{sec:numerical}. We consider the field response $f(\theta,\phi)=\cos^{q}\theta \cos^{q}\phi,$ which yields $|f(\theta,\phi)|^{2}=\cos^{2q}\theta \cos^{2q}\phi.$ Using $d\Omega=\cos\phi\, d\theta\, d\phi$ and the angular domain $\theta \in [-\pi/2,\pi/2]$ and $\phi \in [-\pi/2,\pi/2]$, the normalization integral in \eqref{equ:G_0_definition} can be factorized as
\begin{align}
\int\!\!\!\int |f(\theta,\phi)|^{2}\, d\Omega
&=
\int_{-\pi/2}^{\pi/2}\cos^{2q}\theta\, d\theta
\int_{-\pi/2}^{\pi/2}\cos^{2q+1}\phi\, d\phi.
\label{eq:app_int_factor}
\end{align}

Using the standard integral identity
\begin{equation}
\int_{-\pi/2}^{\pi/2}\cos^{p}x\, dx
=
\sqrt{\pi}\,
\frac{\Gamma\left(\frac{p+1}{2}\right)}{\Gamma\left(\frac{p}{2}+1\right)},
\quad p>-1,
\label{eq:app_cos_int_id}
\end{equation}
we obtain
\begin{equation}
\int_{-\pi/2}^{\pi/2}\cos^{2q}\theta\, d\theta
=
\sqrt{\pi}\,
\frac{\Gamma\left(q+\frac{1}{2}\right)}{\Gamma(q+1)},
\end{equation}
and
\begin{equation}
\int_{-\pi/2}^{\pi/2}\cos^{2q+1}\phi\, d\phi
=
\sqrt{\pi}\,
\frac{\Gamma(q+1)}{\Gamma\left(q+\frac{3}{2}\right)}.
\end{equation}

Therefore,
\begin{equation}
\int\!\!\!\int |f(\theta,\phi)|^{2} d\Omega
=
\pi\,
\frac{\Gamma\left(q+\frac{1}{2}\right)}{\Gamma\left(q+\frac{3}{2}\right)}
=
\frac{2\pi}{2q+1},
\label{eq:app_int_closed}
\end{equation}
where the second equality follows from
$\Gamma\left(q+\frac{3}{2}\right)=\left(q+\frac{1}{2}\right)\Gamma\left(q+\frac{1}{2}\right)$. Finally, substituting \eqref{eq:app_int_closed} into \eqref{equ:G_0_definition} yields
\begin{equation}
G_0
=
4\,
\frac{\Gamma\left(q+\frac{3}{2}\right)}{\Gamma\left(q+\frac{1}{2}\right)}
=
4q+2.
\end{equation}

\bibliographystyle{IEEEtran}
\bibliography{references}

\end{document}

%% file: setting.tex
\usepackage{cite}
\usepackage{amsmath,amssymb,amsfonts,amsthm}
\usepackage{algorithmic}
\usepackage{subcaption}
\usepackage{graphicx}
\usepackage{textcomp}
\usepackage{xcolor}
\usepackage{enumitem}
\usepackage{tabularx,array}

\usepackage{lipsum}
\usepackage{flushend}
\usepackage[normalem]{ulem}
\usepackage{cases}